# Time-Dependent Quantum Monte Carlo for fermions: from Bayesian conditioning to spinor dynamics

Ivan P. Christov

Physics Department, Sofia University, 1164 Sofia, Bulgaria

**Abstract.** The time-dependent quantum Monte Carlo method represents a many-electron state by an ensemble of replicas, in which each electron is a cloud of walkers sampling its density in physical space, one walker and one guide wave per replica, and it replaces the Hartree potential by a conditional interaction built from the walker positions. Here we construct that conditional interaction. Treating the walker as a localization of its electron to a finite resolution, we show that Bayes' theorem and a single empirical substitution yield the Nadaraya–Watson form used heuristically so far, as approximate Bayesian inference, exact as the resolution length vanishes. Because the conditioning removes the gauge freedom which eliminates the orthonormality multipliers in Hartree–Fock, we enforce orthonormality by a term derived from the constraint. We generalize the formulation to spinors, where the Pauli suppression at coincidence is graded by the local spin alignment, with the known two-particle spin equations as its mean-field limit. For two electrons in a double quantum dot in an inhomogeneous magnetic field, against the exact solution, the spinor equations reproduce the non-collinear spin geometry which collinear calculations cannot represent. Determinant replicas recover most of the one-body entanglement entropy missing from a single determinant at strong exchange, with a deficient spin part, which replicas in the form of the spin-coupled pair of valence-bond theory reproduce, together with the exact spin angle and singlet probability in the crossover between the exchange and Zeeman regimes.

**Keywords:** time-dependent quantum Monte Carlo; conditional interaction; Nadaraya–Watson estimator; spinor; entanglement; double quantum dot

## 1. Introduction

The motions of the electrons in atoms, molecules and solids are correlated because of the Coulomb repulsion between them, and the solution of the many-electron Schrödinger equation is therefore a function of 3N variables. If this equation is solved directly on a grid of K nodes in each dimension, the workload is proportional to $K^{3N}$, which concerns both the number of floating point operations and the memory requirements. It is generally believed that such exponential scaling is necessary for any method which pursues an exact solution [1,2]. Two reformulations in physical space are widely used. In the Hartree–Fock approximation the interaction is replaced by the field of an average density, so that the correlation is neglected by construction [3]. In density functional theory the many-body problem is relocated into an exchange–correlation functional of the density [4,5], and its time-dependent extension [6] allows correlated dynamics to be calculated in a numerically tractable way, at the price of a functional whose exact expression is not known. Moreover, it has been shown that not only the density at a given time but its whole history enters the exchange–correlation term in a time-dependent calculation [7].

A second group of methods restores the correlation by expanding the state in configurations. These include the configuration interaction and the coupled-cluster methods [8], the multiconfigurational time-dependent Hartree method [9,10] and its fermionic version [11] for time-dependent problems, and the density matrix renormalization group for low-dimensional systems [12]. Such methods provide controlled accuracy, however the number of configurations which is needed grows rapidly with the number of particles and with the degree of excitation, so that in practice their application to real-time dynamics is limited to systems with few electrons.

A different approach is offered by the quantum Monte Carlo technique, where the state is represented statistically, by a sampled ensemble instead of a basis expansion. In variational Monte Carlo the expectation values of a parametrized trial function are evaluated by Metropolis sampling [13,14]. In diffusion Monte Carlo the Schrödinger equation is propagated in imaginary time as a drift-diffusion-branching process which, for a given nodal surface, converges to the ground state [15–17]. In the auxiliary-field and the continuous-time methods the interaction is replaced by fluctuating one-body fields [18–20]. For ground state energies the accuracy which can be reached at a given cost is difficult to match by any other approach [21–23]. Two features of these methods should be noted. First, the antisymmetry enters through the sign of the sampled amplitude, and the resulting sign problem is in general of exponential difficulty [1]. Second, the sampling is performed in configuration space, where a walker is a point in 3N dimensions rather than a particle in physical space.

Both of these features are important for time-dependent problems. Diffusion Monte Carlo is by its construction an imaginary-time method, since the projection which filters out the ground state has no counterpart in real time, where the propagator is unitary instead of contractive, and the branching which implements that projection has nothing to converge to. A time-dependent calculation therefore needs an object which evolves in physical space and carries the correlation, without the many-body wave function being constructed at any stage. This requirement has become more important with the development of attosecond experiments, where the correlated motion itself is the observable [24].

In a mean-field description each electron experiences only the average charge distributions of its neighbors, not their actual positions. The correlation energy, the mutual avoidance of the electrons and the two-body structure of the pair density, all of which the mean-field theory misses, arise from the difference between the actual and the average positions of the neighbors of a given electron.

Time-Dependent Quantum Monte Carlo (TDQMC) [25–27] restores this information by attaching positions to the electrons. The many-body state is represented by an ensemble of M replicas, and each electron is described by a cloud of M walkers, one per replica, which together sample its density, each carrying its own guide wave. Within a replica *k*, electron *i* therefore has a single guide wave $\varphi_i^k(\mathbf{r},t)$, propagated on a grid, and a single walker $\mathbf{r}_i^k(t)$, moving through space by de Broglie guidance. We call the same function an orbital when its role in the determinant of the replica is meant. The two sectors are coupled as follows: instead of the Hartree potential of the average density, each guide wave feels a conditional potential built from the walkers of the other electrons: the Coulomb field of its neighbors as located, to within a nonlocal length $\sigma$, by their walker positions. Because different replicas have their walkers in different places, the replicas polarize differently, the walkers of interacting electrons avoid each other, and the ensemble acquires a two-body correlation which no single mean-field state can represent. The many-body problem is replaced by a set of coupled one-body equations whose number grows linearly with the number of particles, rather than by a single equation in 3N dimensions, and the cost is polynomial in the number of electrons, the number of replicas and the number of grid points.

The same structure offers a further possibility, which the methods discussed above do not have. The entanglement, the coherence, and the reduced density matrices from which they are calculated are usually reported as single numbers which refer to a chosen bipartition, because the state from which they are calculated is a single object. An ensemble of one-body wave functions is not a single object, since each of its members is labeled by the position of its own walker. Therefore, these quantities acquire a spatial argument and can be resolved in real space, again without constructing the many-body wave function at any stage. In this way the entanglement inside a molecule has been mapped [28], its distribution in lattices which contain structural defects has been located [29], and it has been characterized through the statistics of the one-body ensemble itself [30]. In this respect the particle–wave representation gives access to quantities which the configuration-space methods do not provide, owing to the same feature which makes the real-time problem tractable.

TDQMC changes one term of the time-dependent Hartree–Fock equations, the electron–electron interaction, and only in the equation of motion. The kinetic operator, the nuclear potential and the determinant structure of each replica remain unchanged, and all energies and observables are evaluated with the bare interaction. This picture has been developed and tested, mostly for systems where the interacting electrons have opposite spins. In the present work we extend the formulation to the regime where the exchange is active, namely to spatially overlapping electrons with parallel spins in real time. This requires more than adding a Fock term: we have to determine where the exchange resides, what keeps the orbitals orthogonal, and whether the walkers still follow the guide waves for fermions. None of these questions arises in the singlet benchmarks. In Section 5 we test the spinor equations on two electrons in a double quantum dot in an inhomogeneous magnetic field, against the exact solution, and Section 6 summarizes.

Our primary aim is to construct the conditional interaction from a conditional model and Bayes' theorem, which gives the effective-potential construction of the earlier work a statistical interpretation and its resolution length a meaning which is independent of any fit. The spinor formulation generalizes the same construction, and the local screening representation of Section 2.3 is an optional numerical alternative to the explicit exchange. In this work we simulate the imaginary-time propagation only, in which the walkers sample the converged density and, for determinant replicas, are then held fixed.

Compared with the earlier formulations of TDQMC [25–27], the present work adds four elements. The conditional interaction, which has been used in the Nadaraya–Watson form with a width fixed by energy minimization, is constructed here from a conditional model and Bayes' theorem, and we show that at a finite width this construction is approximate Bayesian, becoming exact Bayes as the width vanishes (Section 2.2). For parallel spins, where orthonormality has been restored by explicit orthogonalization, we derive the projection term which the conditioning requires and the source which this term and the exchange generate in the continuity equation of the guide-wave density (Sections 2.4 and 3). The formulation is extended to spinors (Section 4). Finally, the spinor equations are compared with the exact solution for two electrons whose spins are not collinear, at the level of the one-body density matrix and of the singlet probability (Section 5). The method is not variational, since the conditional potential enters the equation of motion only (Section 2.2, P3).

## 2. From the mean field to the conditional interaction

### 2.1 The two retained surfaces of the replica pair density

The Dirac–Frenkel action [31,32] of replica $k$, evaluated on its determinant, produces the pair energy as a sum over two retained surfaces of the two-body density $\rho_2^k$, in which $\mathbf{a}$ and $\mathbf{b}$ are the positions of the two electrons of each pair. In the self-interaction-free convention forced in Appendix A, these are the **direct (charge) surface,** which is the product of orbital densities $\left|\varphi_m^k(\mathbf{a})\right|^2\left|\varphi_n^k(\mathbf{b})\right|^2$ summed over pairs $m \neq n$ (1a), and the **exchange (coherence) surface**, entering with weight $\delta_{s_m s_n}$, where $s_m$ is the spin label of orbital $m$, (1b),

$$\rho_2^k(\mathbf{a},\mathbf{b}) = \sum_{m\neq n}\left(\left|\varphi_m^k(\mathbf{a})\right|^2\left|\varphi_n^k(\mathbf{b})\right|^2 - \delta_{s_m s_n} w_{mn}^k(\mathbf{a},\mathbf{b})\right), \tag{1a}$$

$$w_{mn}^k(\mathbf{a},\mathbf{b}) = \varphi_m^k(\mathbf{a})\varphi_m^{k*}(\mathbf{b})\varphi_n^k(\mathbf{b})\varphi_n^{k*}(\mathbf{a}). \tag{1b}$$

The excluded $m = n$ (self-pair) surface is discarded identically from both sectors, and the remaining surfaces are those the single-determinant ansatz cannot represent. The two retained surfaces differ in nature: the first is a probability density, the second a coherence. They give the two interaction terms of the equations of motion. For each electron $j$ the method has one parameter, the nonlocal length $\sigma_j$, which acts on the direct surfaces of electron $j$ and produces the correlation (Section 2.2) while the exchange surface is not conditioned (Section 2.3).

In the Hartree–Fock limit (all replicas the same) both surfaces enter the dynamics through a single equation, which for each replica $k$ and $N$ orthonormal orbitals $\varphi_i^k(\mathbf{r},t)$ reads [3]

$$i\hbar\partial_t\varphi_i^k = \left(h + J_i^k - \hat{K}_i^k\right)\varphi_i^k. \tag{2}$$

in which $h$ holds the kinetic energy of an electron of mass $m_\mathrm{e}$ and the external potential $V_{en}$ of the nuclei, $h = -\frac{\hbar^2}{2m_\mathrm{e}}\nabla^2 + V_{en}(\mathbf{r})$, while the Hartree operator $J_i^k$ and the exchange operator $\hat{K}_i^k$ are the functional derivatives, with respect to $\varphi_i^{k*}$, of the direct and of the exchange surface respectively. They are thus the operator forms of the charge surface and of the coherence surface, and both are written out explicitly in Section 2.4, where the orthonormality constraint is carried through as well.

### 2.2 From the Hartree potential to the conditional interaction: definitions and construction

**Setup and sampling assumption.** Each electron $j$ carries one walker per replica, $\mathbf{r}_j^l$, $l = 1,\dots,M$. Throughout, $k$ denotes the replica whose equation of motion is being written (the conditioning replica) and $l$ is the summation index over the pooled ensemble. The construction rests on one sampling assumption:

**(A0)** At each time $t$, the walker $\mathbf{r}_j^l(t)$ of replica $l$ is distributed according to the density $\left|\varphi_j^l(\mathbf{r},t)\right|^2$ of its own guide wave, and the pooled ensemble $\left\{\mathbf{r}_j^l(t)\right\}_{l=1}^M$ remains a representative sample across replicas.

In imaginary time the assumption A0 expresses the stationarity of the drift-diffusion process (Section 3). In real time it expresses the equivariance of the de Broglie guidance, which is exact for the local terms of the equation of motion. The exchange operator and, as is shown in Section 2.4, the projection term generate a source in the continuity equation, and the sampling bias which results from it is treated in Section 3.

Under A0 the pooled ensemble is a sample of size $M$ from the mixture density

$$n_j(\mathbf{r},t) = \frac{1}{M}\sum_{l=1}^{M}\left|\varphi_j^l(\mathbf{r},t)\right|^2. \tag{3}$$

At the replica-symmetric point (identical orbitals: $\sigma_j \to \infty$ fixed point) the mixture coincides with each replica's density; after replica symmetry breaking it does not.

**The definition of the conditioning.**

**(P1)** In the equation of motion of replica $k$, the interaction of electron $i$ with electron $j$ is conditioned on replica $k$'s own walker $\mathbf{r}_j^k$, treated as a $\sigma_j$-resolution proxy for the position $\mathbf{R}_j$ of electron $j$. Formally, the conditional probability is

$$p\left(\mathbf{r}_j^k \mid \mathbf{R}_j = \mathbf{r}\right) = K_{\sigma_j}\left(\mathbf{r}_j^k - \mathbf{r}\right), \tag{4}$$

with $K_{\sigma_j}$ a normalized, symmetric kernel of width $\sigma_j$; for the Gaussian kernel, $K_{\sigma_j}(\mathbf{u}) = \left(2\pi\sigma_j^2\right)^{-d/2}\exp\left(-|\mathbf{u}|^2/2\sigma_j^2\right)$ in $d$ dimensions.

P1 defines the point where the corpuscular information of replica $k$ enters the wave sector, and it fixes the physical meaning of $\sigma_j$ at the outset: $\sigma_j$ is the spatial resolution with which the walker localizes electron $j$, the nonlocality length of the conditional interaction.

**Bayes step.** Here we use the standard terminology of Bayesian statistics (see, e.g., [33]) where $n_j$ is called the prior, (4) is the likelihood such that for the prior $\mathbf{R}_j \sim n_j$ and the conditional model of P1, Bayes' theorem gives the posterior density $p_j^k(\mathbf{r})$ of $\mathbf{R}_j$ (the posterior). As in approximate Bayesian computation [34,35], where the likelihood is replaced by its convolution with a kernel of finite width, this posterior is exact for the conditional model P1, in which the walker is the position of its electron plus a kernel-distributed error [34], and it becomes the exact posterior as $\sigma_j \to 0$ [35]. The posterior reads

$$p_j^k(\mathbf{r}) = \frac{K_{\sigma_j}\left(\mathbf{r}-\mathbf{r}_j^k\right) n_j(\mathbf{r})}{\int K_{\sigma_j}\left(\mathbf{r}'-\mathbf{r}_j^k\right) n_j(\mathbf{r}')\, d\mathbf{r}'}. \tag{5}$$

The population-level conditional potential for electron $i$ in replica $k$ then defined, with $V_{ee}$ the electron–electron interaction potential:

$$V^k(\mathbf{r}_i, t) = \sum_{j \neq i} \int V_{\text{ee}}(\mathbf{r}_i - \mathbf{r})\, p_j^k(\mathbf{r})\, d\mathbf{r}. \tag{6}$$

**Empirical plug-in (the statistical step).** The mixture $n_j$ is not available in closed form; what is available under A0 is the pooled walker sample. Replacing $n_j$ by the empirical measure $n_j(\mathbf{r}) = M^{-1} \sum_l \delta\left(\mathbf{r}-\mathbf{r}_j^l\right)$ in both numerator and denominator of (5)–(6) gives,

$$V_{\text{eff}}^k(\mathbf{r}_i, t) = \sum_{j \neq i} \frac{\sum_l V_{\text{ee}}\left(\mathbf{r}_i - \mathbf{r}_j^l\right) K_{\sigma_j}\left(\mathbf{r}_j^l - \mathbf{r}_j^k\right)}{\sum_l K_{\sigma_j}\left(\mathbf{r}_j^l - \mathbf{r}_j^k\right)}, \tag{7}$$

which is the Nadaraya–Watson form [36,37], the kernel-weighted posterior mean of approximate Bayesian computation with the pooled walkers, the conditioning walker included, as the samples of the prior and the conditioning walker as the observation. No kernel density estimate and no sharply-peaked-kernel approximation enters: the substitution is exact for the empirical measure [27]. The only statistical statement required is the law of large numbers at fixed $\sigma_j$: given A0 and weak correlation between the walkers of different replicas, $V_{\text{eff}}^k \to V^k$ as $M \to \infty$, with fluctuations of order $M_{\text{eff}}^{-1/2}$ and, since $V_{\text{eff}}^k$ is a ratio of Monte Carlo averages, an additional ratio-estimator bias of order $M_{\text{eff}}^{-1}$, where

$$M_{\text{eff}}^k = \frac{\left(\sum_l w_l\right)^2}{\sum_l w_l^2}, w_l = K_{\sigma_j}\left(\mathbf{r}_j^l - \mathbf{r}_j^k\right) \tag{8}$$

is the effective number of walkers inside the kernel window.

**Limits.** Three consequences follow directly.

**1.** $\sigma_j \to \infty$ for all $j$: the weights become uniform and

$$V_{\text{eff}}^{k} \to \sum_{j \neq i} \frac{1}{M} \sum_{l} V_{\text{ee}}\left(\mathbf{r}_i - \mathbf{r}_j^{l}\right), \tag{9}$$

the Monte Carlo estimate of $\sum_{j \neq i} \int V_{\text{ee}}\left(\mathbf{r}_i - \mathbf{r}\right) n_j\left(\mathbf{r}\right) d\mathbf{r}$, the Hartree potential of the mixture density, identical for all replicas. TDHF is recovered with the limits taken in this order: $\sigma_j \to \infty$ first makes the equation of motion replica-independent, so identically initialized replicas remain identical and $n_j = \left|\varphi_j\right|^2$; then $M \to \infty$ converges the walker average to the Hartree integral.

**2.** $\sigma_j \to 0$: the kernel weights concentrate on the $l = k$ walker and $V_{\text{eff}}^{k} \to \sum_{j \neq i} V_{\text{ee}}\left(\mathbf{r}_i - \mathbf{r}_j^{k}\right)$, the bare pair potential evaluated at the instantaneous walker position, which is the pair limit, singular for the bare Coulomb interaction [26].

**3.** Finite $\sigma_j$: expanding the posterior (5) about $\mathbf{r}_j^{k}$ gives, per pair, for the Gaussian kernel of P1,

$$\int V_{\text{ee}}\left(\mathbf{r}_i - \mathbf{r}\right) p_j^{k}\left(\mathbf{r}\right) d\mathbf{r} = V_{\text{ee}}\left(\mathbf{r}_i - \mathbf{r}_j^{k}\right) - \sigma_j^2 \nabla V_{\text{ee}}\left(\mathbf{r}_i - \mathbf{r}_j^{k}\right) \cdot \nabla \ln n_j\left(\mathbf{r}_j^{k}\right) + \frac{\sigma_j^2}{2} \nabla^2 V_{\text{ee}}\left(\mathbf{r}_i - \mathbf{r}_j^{k}\right) + O\left(\sigma_j^4\right). \tag{10}$$

The two terms in the expansion (10) carry different coefficients because they describe different things. The first is a displacement: weighting the kernel by $n_j$ moves the $\sigma_j$-sized region within which the walker locates electron $j$ off the walker itself and toward higher density, by $\sigma_j^2 \nabla \ln n_j$. The second is a spread: the potential is averaged across that region instead of being taken at a single point. Expanding $V_{ee}$ to second order about $\boldsymbol{r}_j^{k}$ therefore returns $\sigma_j^2$ on the displacement and $\frac{\sigma_j^2}{2}$ on the Laplacian. The factor of two between them is not a property of the Gaussian. The $O\left(\sigma_j^2\right)$ terms belong to the model: they describe the finite-resolution conditioning that interpolates between the pair limit and the mean field. (For the bare Coulomb interaction $\nabla^2 V_{\text{ee}}$ is a contact term and the expansion is formal; it is regular for soft-Coulomb interactions.)

**The remaining definitions of the direct sector.**

**(P2)** The kernel widths are fixed by energy minimization [38,39]: each $\sigma_j$ is set to the value at which the system energy $E\left(\sigma_j\right)$ is lowest. The numerical observation which motivates the criterion is a well-defined minimum lying below the mean-field energy [27,40].

**(P3)** The replacement $J_i^{k} \to V_{\text{eff}}^{k}$ is made in the equation of motion only. The energy functional and all observables use the bare interaction $V_{en}$ and $V_{ee}$; $V_{\text{eff}}^{k}$ never appears in the energy. The bare-interaction energy is a diagnostic, not a constant of the motion: its drift in real time measures the error of the conditional dynamics, and the minimum in (P2) is a prescription for the method rather than a variational bound.

### 2.3 Exchange in the particle–wave representation: defining the partition

The equations of motion contain two interaction terms of different origin: the conditional potential $V_{\mathrm{eff}}^{k}$, built from walker statistics through the Bayes–NW chain, and the exchange operator, inherited from the determinant structure of the replica. Here we show how the two are connected and why the exchange term cannot be given the same statistical construction.

The Dirac–Frenkel action of replica $k$ produces the pair energy of the determinant as the sum of the two retained surfaces of $\rho_2^k$ (Section 2.1),

$$E_{\mathrm{ee}}^{k} = E_{\mathrm{H}}^{k} + E_{\mathrm{X}}^{k}, \tag{11}$$

where $E_H^k$ and $E_X^k$ are the Hartree and the exchange energy of the replica, Eqs. (A.1)–(A.2), and its functional derivative delivers $J_i^k - \hat{K}_i^k$ together (Appendix A, Step 2). Which of the two is conditioned is therefore decided by a single criterion, whether the surface can be represented by a probability density: the direct surface is a positive, normalized pair density and receives the replacement $J_i^k \to V_{\mathrm{eff}}^{k}$ of P1–P3, while the coherence surface does not.

The exchange surface, taken on its own, admits no positive statistical representation, for the following reason. The Bayes–NW construction operates only on probability measures: nonnegative, normalizable densities of walker positions. The pair exchange surface $w_{mn}^{k}(\mathbf{a},\mathbf{b})$ of (1b) is complex in general and sign-indefinite even for real orbitals. Moreover, its total "mass" vanishes:

$$\iint w_{mn}^{k}(\mathbf{a},\mathbf{b})\, d\mathbf{a}\, d\mathbf{b} = \left| < \varphi_m^k \mid \varphi_n^k > \right|^2 = 0, \quad (m \neq n), \tag{12}$$

by the orthogonality of the determinant orbitals. A signed measure of zero total "mass" admits no importance-sampling normalization, so any conditional (NW-type) estimator built on it has a denominator which fluctuates about zero, and conditioning the exchange on walker positions is undefined. The obstruction is narrower than the fermion sign problem: the determinant pair density itself is nonnegative and can be sampled with positive weights [13,14], but the one-body conditioning of P1 could carry the exchange only as the separate surface (1b), which has no positive representation. Leaving the exchange to the wave sector therefore keeps the walker dynamics positive.

**(P4)** Corpuscular information conditions the charge sector of $\rho_2^k$ only. The coherence sector admits no positive, normalized walker representation and is propagated variationally, per replica, from the determinant structure. Exchange is accounted for once, in the wave sector, and the particle sector is exchange-free. P4 is a definition of the method, on the same footing as P1–P3. At the ensemble level the accounting is exact: the two-body density of the replica mixture is the replica average of the determinant two-body densities, so the ensemble-averaged exchange surface $M^{-1}\sum_k w_{ij}^k$ is the exchange component of the mixture's $\rho_2$. One consequence of P4 is immediate. At $\mathbf{a}=\mathbf{b}$ and equal spins the direct and exchange surfaces of any replica cancel identically,

$$\left|\varphi_i^k(\mathbf{a})\right|^2\left|\varphi_j^k(\mathbf{a})\right|^2-w_{ij}^k(\mathbf{a},\mathbf{a})=0, \tag{13}$$

an algebraic identity which holds pointwise and does not require orthogonality of the orbitals, so that the effective same-spin pair density of the energy estimator vanishes at coincidence and the exact determinant-level Pauli suppression is reproduced replica by replica. Its spinor form, graded by the local spin alignment, is Eq. (32) of Section 4. This cancellation concerns the pair density of the determinant; the joint distribution of the walkers has no node, since walkers of different electrons may coincide.

Thus, P4 forbids one operation only, the sampling of the exchange surface by walkers, and two representations of that surface remain. The first representation is the Fock operator of the wave sector, which is exact, and it is the natural choice in imaginary time, because the walkers of the imaginary-time stage (Section 3) sample the instantaneous guide-wave density whatever operator propagates it on the grid; the nonlocality of the exchange and projection terms is therefore invisible to the sampling, and since the imaginary-time orbitals are real no source is generated. The second representation is a screening factor multiplying the interaction inside the charge sector [41],

$$V_{\mathrm{ee}}(\boldsymbol{r}-\boldsymbol{r}') \;\rightarrow\; V_{\mathrm{ee}}(\boldsymbol{r}-\boldsymbol{r}')\ \operatorname{erf}\left(\frac{|\boldsymbol{r}-\boldsymbol{r}'|}{r_{\mathrm{s}}}\right), \tag{14}$$

which models the exchange hole instead of carrying its coherence. Such a factor does not sample the exchange surface and therefore leaves P4 intact, since it folds the effect of the hole into the local potential which the walkers already experience. Its advantage is the exact equivariance, because a real local potential contributes nothing to the continuity balance of the guide-wave density; the screened form therefore generates no source, and the correction of walkers trajectories (Section 3) is not needed. Its cost is that the representation is approximate and that it carries a length scale of its own. The two representations must not act together, since both remove the same hole. The screening length $r_{\mathrm{s}}$ is not a free parameter: requiring the screened direct term to reproduce the exchange integral of the same reference state,

$$\iint V_{\mathrm{ee}}(\boldsymbol{r}-\boldsymbol{r}')\ n_1(\boldsymbol{r})\ n_2(\boldsymbol{r}')\left[1-\operatorname{erf}\left(\frac{|\boldsymbol{r}-\boldsymbol{r}'|}{r_{\mathrm{s}}}\right)\right]\ \mathrm{d}\boldsymbol{r}\ \mathrm{d}\boldsymbol{r}'=K_{12}, \tag{15}$$

where $n_1$ and $n_2$ are the densities of the two orbitals of the pair and $K_{12}$ is their exchange integral (A.2), determines it without fitting. Since the exchange hole does not change appreciably between the correlated regime and the Hartree–Fock one, the screening length is calibrated in practice on the TDHF reference state [41].

### 2.4 Derivation of the exchange-involved TDQMC equations (see Appendix A)

As shown in Appendix A, the TDQMC equation of motion of replica $k$ reads

$$i\hbar\,\partial_t\varphi_i^k=F_i^k\,\varphi_i^k-\sum_j\lambda_{ij}^k\,\varphi_j^k. \tag{16}$$

in which the generator of orbital $i$ in replica $k$ is

$$F_i^k = h + V_{\text{eff}}^k(\mathbf{r}_i) - \hat{K}_i^k, \tag{17}$$

the one-body operator carrying the conditional interaction of Section 2.2 in place of the Hartree term, with the exchange operator $\hat{K}_i^k$ retained.

The $\lambda_{ij}^k$ in (16) are the Lagrange multipliers of the orthonormality constraints $<\varphi_i^k \mid \varphi_j^k>=\delta_{ij}$, one for each constrained pair. They are needed because $V_{\text{eff}}^k$ is built from the walkers of the electrons other than $i$, and thus the generators of different orbitals of the same replica differ. They are not free parameters: they are fixed by the requirement that orthonormality, which holds at one instant, be preserved by the evolution. Imposing that requirement gives the canonical choice $\lambda_{ij}^k =<\varphi_j^k \mid F_i^k\, \varphi_i^k>$, and substituting it into (16) collapses the sum into the projection,

$$i\hbar\,\partial_t \varphi_i^k = F_i^k\, \varphi_i^k - \sum_{j\neq i} \varphi_j^k < \varphi_j^k \mid F_i^k\, \varphi_i^k > = \left(1 - \sum_{j\neq i} \left|\varphi_j^k ><\varphi_j^k\right|\right) F_i^k\, \varphi_i^k, \tag{18}$$

which serves as the exchange-involved TDQMC equations of motion in orthonormality-preserving form. Equation (18) does not resemble the standard time-dependent Hartree–Fock equation (2) because (2) is the same problem in gauge-fixed form, in which the multipliers are Hermitian and transformed away. Setting the kernel width to infinity returns (2), the reference calculation against which the finite-width results are measured.

**(P5)** The multiplier gauge is fixed by the canonical (minimal) choice (A.14), $\lambda_{ii}^k = 0$. The constraint condition (A.13) fixes only the anti-Hermitian combination $\lambda_{ji}^k - \lambda_{ij}^{k*}$; adding any Hermitian matrix yields another member which enforces the same orthonormality and mixes the occupied orbitals without changing the determinant. At finite $\sigma_j$, however, such a mixing changes the subsequent dynamics, since the conditional potential is attached to the individual orbitals through their walkers. Two discrete members are used in practice: per-step symmetric (Löwdin) orthonormalization [42] of the same-spin orbitals, and per-step sequential Gram–Schmidt, which removes the whole overlap error from the later orbital of each pair. Both differ from the canonical member (18) only by rotations of the size of the per-step overlap error, and the three members therefore agree to integrator order for determinant-level observables, although not at the level of individual walker trajectories, where the guidance velocity is built from the individual orbitals. Which member is used is therefore part of the specification of a TDQMC calculation. All three reduce to the standard TDHF gauge as $\sigma_j \to \infty$, where the projection term disappears altogether, as derived at the end of Appendix A, and whichever is chosen, the projection term adds nothing to the instantaneous energy, since $<\varphi_i^k \mid \sum_{j\neq i} \lambda_{ij}^k \varphi_j^k >= 0$ under orthonormality.

The projection term exists only for fermions. A bosonic replica is a Hartree product, whose ansatz imposes no mutual orthogonality, so that there is no off-diagonal constraint and no projection term. The multipliers are absent from the published equations because $V_{\text{eff}}$ was substituted into the already gauge-fixed TDHF equation, and that substitution removes the common-generator property on which the gauge fixing relied. In practice orthonormality has been restored explicitly, by Gram–Schmidt for parallel spins [41], and no published result is affected. Re-orthogonalization at each time step is the discrete realization of the projection term in the sense of P5: a calculation which performs it already solves (18) in the corresponding gauge, to integrator order for determinant-level observables. For spinors the projection spans all orbital pairs.

Although TDQMC uses guidance equations formally similar to those of Bohmian mechanics [43,44], it is neither a Bohmian theory nor a reformulation of one: the Bohmian guiding wave is a many-body function on configuration space, whereas here the guiding objects are one-body wave functions in physical space, so that the walker trajectories are generated by physical-space dynamics. The walkers are point objects guided by the de Broglie relation, and they are statistical sampling objects rather than the constituents of a single physically realized configuration, while the guide waves are auxiliary one-body fields introduced in order to represent correlated many-body dynamics without constructing the full many-body wave function.

## 3. Walker dynamics

The walker ensemble enters in two dynamical regimes which have a different status. In imaginary time the walkers are samplers, in the sense that they track a density which is propagated on the grid by the guide-wave equation, and their dynamics is chosen for its stationary distribution. In real time they are propagators, since they transport the initial samples forward, and their correctness then depends on a continuity property of the guide-wave equation itself.

**Stage 1: Imaginary-time relaxation.** Setting $\tau = it$, Eq. (18) for electron $i$ of replica $k$ becomes

$$-\hbar\,\partial_\tau \varphi_i^k = \left(1 - \sum_{j\neq i} \left|\varphi_j^k \right> \left< \varphi_j^k \right|\right)\left(h + V_{\text{eff}}^k(\mathbf{r}_i) - E_{\text{ref}} - \hat{K}_i^k\right)\varphi_i^k, \tag{19}$$

where $E_{\text{ref}}$ compensates the norm contraction; in practice the guide wave is renormalized after each step, which is equivalent and changes nothing below, because the walker dynamics depends on the density only through $\nabla \ln\left|\varphi_i^k\right|^2$, invariant under normalization. As $\tau \to \infty$ the guide waves approach the lowest self-consistent stationary state of the nonlinear, walker-dependent operator; since $V_{\text{eff}}^k$ fluctuates with the walker positions, the limit is a *stochastic* fixed point, and stationarity is diagnosed through the energy estimator (P2).

**Walker dynamics.** The walkers evolve by the overdamped Langevin process

$$d\mathbf{r}_j^k = \mathbf{u}_j^k\left(\mathbf{r}_j^k,\tau\right)d\tau + \sqrt{2D\,d\tau}\;\xi, \qquad \mathbf{u}_j^k = D\nabla\ln\left|\varphi_j^k\right|^2 = \frac{\hbar}{m_{\text{e}}}\frac{\nabla\varphi_j^k}{\varphi_j^k}, \qquad D = \frac{\hbar}{2m_{\text{e}}}, \tag{20}$$

with $\xi$ Gaussian, zero mean, unit covariance. In (20) $\mathbf{u}_j^k$ is the osmotic velocity of stochastic mechanics [45–48], which pushes the walkers up the density gradient. Let $f(\mathbf{r},\tau)$ denote the walker density. It obeys the Fokker–Planck equation

$$\partial_\tau f = -\nabla\cdot(\mathbf{u}f) + D\nabla^2 f, \tag{21}$$

whose stationary solution for frozen $\varphi$ is $f \propto |\varphi|^2$ for *any* $D > 0$. For a real solution $\varphi$ of the imaginary-time propagation, substituting $\rho = \varphi^2$ into the Fokker–Planck transport terms gives an identity, $-\nabla \cdot (\mathbf{u}\rho) + D\nabla^2 \rho = 0$, so the drift–diffusion process leaves the instantaneous target invariant. The actual $\tau$-evolution of $\rho$ is carried entirely by the *reaction* term,

$$\partial_\tau \rho = -\frac{2}{\hbar}\left(E_L(\mathbf{r},\tau) - E_{\text{ref}}\right)\rho, \qquad E_L = \frac{H_{\text{eff}}^k \varphi_i^k}{\varphi_i^k}, \tag{22}$$

where $H_{\text{eff}}^k$ collects all terms of the equation of motion, *including the projection*. In Diffusion Monte Carlo this term is implemented as branching or weights [17,21,22]; in TDQMC no branching is needed, because $\rho$ is not reconstructed from the walkers, since it is propagated on the grid, and the Langevin process only re-equilibrates the walkers to a slowly moving target. The validity condition is quasi-stationarity: the Langevin mixing time over the length scale on which $\nabla \ln \rho$ varies must be short compared with the relaxation time of the guide waves. Residual finite-$d\tau$ discretization bias is removed by a Metropolis–Hastings acceptance step with target $|\varphi|^2$ (Metropolis-adjusted Langevin) [49]. In the calculations of Section 5 the walkers of the determinant replicas are moved by the Metropolis algorithm with the same target during the first relaxation pass only, and the guide waves then relax further around fixed walker positions, each replica in the conditional potential of a fixed position of the partner walker. These steps bring the guide waves to their optimal form for the walkers which start the real-time propagation, which is important for its stability, and the quasi-stationarity condition is then met trivially, since the walker set remains a fixed sample of the replica-averaged density $n_j$ of (3) as long as $n_j$ does not change. With appropriate parameters and number of steps they change the energy by less than 1.5% (Appendix C.1).

**Stage 2: Real-time propagation.** At $t = 0$ the perturbation is switched on and the guide waves obey (18). The walkers move by the guidance law

$$\frac{d\mathbf{r}_j^k}{dt} = \mathbf{v}_j^k\left(\mathbf{r}_j^k, t\right), \qquad \mathbf{v} = \frac{\mathbf{j}}{|\varphi|^2}, \qquad \mathbf{j} = \frac{\hbar}{m_\text{e}} \operatorname{Im}\left(\varphi^* \nabla \varphi\right). \tag{23}$$

**Equivariance and its hypothesis.** Under pure transport, $\partial_t f + \nabla \cdot (\mathbf{v} f) = 0$, the ratio $f / |\varphi|^2$ is constant along trajectories; hence if $f(\cdot,0) = |\varphi(\cdot,0)|^2$ then $f(\cdot,t) = |\varphi(\cdot,t)|^2$ for all $t$, *provided* $|\varphi|^2$ itself obeys the same continuity equation, $\partial_t |\varphi|^2 + \nabla \cdot \mathbf{j} = 0$. (With one walker per replica, "distributed as" is a statement in probability across the replica ensemble: each walker remains a correct sample of its own replica's density, which is A0.) Continuity holds for every *local* term of (18): the kinetic term, $V_\text{en}$, $V_\text{eff}^k$ (real and local by construction), and, in the spinor case, the Zeeman term (Section 4), which is local and Hermitian in spin space and contributes nothing to $\partial_t \chi^\dagger \chi$. It fails for the exchange and projection terms.

Writing (18) as $i\hbar \partial_t \varphi_i^k = H_\text{loc} \varphi_i^k - \hat{K}_i^k \varphi_i^k - \sum_{j \neq i} \lambda_{ij}^k \varphi_j^k$, where $H_{loc} = h + V_{eff}^k(\mathbf{r}_i)$ collects the local terms, direct computation gives

$$\partial_t \left|\varphi_i^k\right|^2 + \nabla\cdot\mathbf{j}_i^k = s_i^k(\mathbf{r},t), \qquad s_i^k = -\frac{2}{\hbar}\mathrm{Im}\left[\varphi_i^{k*}(\mathbf{r})\left(\hat{K}_i^k\varphi_i^k + \sum_{j\neq i}\lambda_{ij}^k\varphi_j^k\right)(\mathbf{r})\right], \tag{24}$$

in which the first term of the bracket is the exchange part of the source and the second is the projection part. Explicitly, the exchange part reads

$$s_{i,\mathrm{X}}^k(\mathbf{r},t) = -\frac{2}{\hbar}\sum_{j\neq i}\delta_{s_is_j}\,\mathrm{Im}\left[\varphi_i^{k*}(\mathbf{r})c_{ij}^k(\mathbf{r})\varphi_j^k(\mathbf{r})\right], \qquad c_{ij}^k(\mathbf{r}) = \int d\mathbf{r}' V_{\mathrm{ee}}(\mathbf{r}-\mathbf{r}')\varphi_j^{k*}(\mathbf{r}')\varphi_i^k(\mathbf{r}'). \tag{25}$$

The total "mass" of the source vanishes, $\int s_i^k\, d\mathbf{r} = 0$ for two independent reasons: the exchange part integrates to $-\frac{2}{\hbar}\mathrm{Im}<\varphi_i^k\,|\,\hat{K}_i^k\varphi_i^k>=0$ by Hermiticity of the Fock operator, while the projection part integrates to $-\frac{2}{\hbar}\mathrm{Im}\sum_{j\neq i}\lambda_{ij}^k<\varphi_i^k\,|\,\varphi_j^k>=0$ by orthogonality. Exchange conserves the norm but transports density nonlocally; the projection redistributes it but is by construction orthogonal to the orbital it acts on. In neither case does a convective velocity field appear in the equation to track the transport. Deterministically guided walkers therefore drift away from $\left|\varphi_i^k\right|^2$ at the rate set by $s_i^k$, and $V_{\mathrm{eff}}^k$ inherits the bias.

**When the source vanishes.** $s_i^k \equiv 0$ in three situations: (i) for pairs with $\delta_{s_is_j} = 0$ , in particular for the two-electron singlet, where exchange part and projection part vanish together, which is why 1D para-helium benchmarks do not show this effect; (ii) whenever all same-spin orbitals are effectively real, as at $t = 0$ after Stage 1 in the collinear case, so the source switches on only as dynamical phase differences develop; (iii) when same-spin orbitals do not overlap spatially, and then both $c_{ij}^k \to 0$ and $\lambda_{ij}^k \to 0$.

The walker drift for fermions is corrected by restoring the continuity balance directly. If we add a correction field $\mathbf{v}_\mathrm{X}$ to walker's velocity satisfying

$$\nabla\cdot\left(\left|\varphi_i^k\right|^2\mathbf{v}_\mathrm{X}\right) = -s_i^k; \tag{26}$$

the corrected guidance $\mathbf{v}+\mathbf{v}_\mathrm{X}$ then restores exact equivariance: $\partial_t|\varphi|^2 + \nabla\cdot\left[|\varphi|^2(\mathbf{v}+\mathbf{v}_\mathrm{X})\right] = s + \nabla\cdot\left(|\varphi|^2\mathbf{v}_\mathrm{X}\right) = 0$. Solvability is guaranteed by $\int s_i^k\, d\mathbf{r} = 0$, which follows from the Hermiticity of the Fock operator and the orthonormality of the replica orbitals. In one dimension the solution is closed-form,

$$v_\mathrm{X}(x,t) = -\frac{1}{\left|\varphi_i^k(x,t)\right|^2}\int_{-\infty}^{x} s_i^k(x',t)\,dx', \tag{27}$$

a single quadrature per guide wave per time step. Since the total source vanishes, the flux $|\varphi_i^k|^2\; v_\mathrm{X}$ vanishes at both ends of the grid, so that the correction redistributes the walkers inside the grid and carries none across its ends. The velocity $v_\mathrm{X}$ itself is this flux divided by $|\varphi_i^k|^2$, and it is not

bounded where the density is small, in the tails and at the nodes of the guide wave, where the flux does not vanish in general. There (27) has to be regularized, for instance by the soft division $|\varphi_i^k|^2/(|\varphi_i^k|^4+\varepsilon^2)$ with $\varepsilon$ small compared with the maximum of $|\varphi_i^k|^2$, together with a bound on $|v_X|$, and the equivariance is then exact only where the density is not small. In higher dimensions one solves a Poisson-type problem, $|\varphi|^2\,\mathbf{v}_X=\nabla\Lambda$ with $\nabla^2\Lambda=-s$, modulo the usual divergence-free gauge freedom. The corrected guidance is not used in the imaginary-time calculations of Section 5, and its test belongs to the real-time calculations which will be reported elsewhere.

TDQMC starts a real-time propagation after the initial state of the system is well established and the conditioning of P1 confines the influence of one particle on another to a neighborhood of width $\sigma_j$ [27,50]. TDQMC thereby gives up the instantaneous influence of a local change throughout configuration space, which allows a low-order polynomial scaling of the cost per time step, unlike the multiconfigurational methods [9–11]. A calculation which begins from an uncorrelated ensemble and switches the interaction on abruptly is therefore outside the scope of the method. Such sudden switches can be handled by the exact solution of TDSE or by other methods [51] where a change at one point of configuration space is felt throughout it at once (nonlocal causality effect) but at much higher cost.

## 4. Spinor extension

The construction of Sections 2.2–2.4 carries a hidden kinematic restriction: each orbital $\varphi_i^k$ was assigned a *fixed* spin label $s_i$, and the exchange operator (A.5) was contracted with $\delta_{s_is_j}$. This is the collinear assumption, in which every electron is a spin eigenstate along a common axis at all times. It fails where spin becomes dynamical: transverse magnetic-field components, spin–orbit terms [52,53], or any initial state whose spinors are not parallel. The generalization requires two substitutions: the orbital is promoted to a two-component spinor [54,55], and the Kronecker contraction of the exchange term is promoted to a spin trace. The Bayes–NW chain, the partition definition, the projected equation of motion, the source term and its correction then remain valid with the spinor inner product in place of the scalar one, and two steps change their form, the graded no-double-counting condition and the projection over all orbital pairs. We verify each step below; the derivation is given in Appendix B.

**Kinematics.** Each orbital of replica k becomes a two-component spinor,

$$\chi_m^k(\mathbf{r},t)=\begin{pmatrix}\varphi_{m\uparrow}^k(\mathbf{r},t)\\ \varphi_{m\downarrow}^k(\mathbf{r},t)\end{pmatrix}, \tag{28}$$

with the orthonormality constraint now read in the full spinor inner product, $<\chi_i^k\,|\,\chi_j^k>=\int\chi_i^{k\dagger}\chi_j^k(\mathbf{r})\,d\mathbf{r}=\delta_{ij}$. No spin labels $s_i$ exist any longer; spin resides in the spatial dependence of the two components. The orbital overlap of the replica is the 2×2 matrix in spin space

$$\Gamma^k(\mathbf{r};\mathbf{r}')=\sum_m\chi_m^k(\mathbf{r})\,\chi_m^{k\dagger}(\mathbf{r}'), \tag{29}$$

and the total charge density is its trace over spin, $n^k(\mathbf{r})=\mathrm{Tr}_s\,\Gamma^k(\mathbf{r};\mathbf{r})=\sum_m\chi_m^{k\dagger}\chi_m^k(\mathbf{r})$.

The determinant structure survives the promotion, since the derivation of Appendix A used only the permutation symmetry and the orthonormality, both of which remain intact. The diagonal of the two-body density (1a) becomes

$$\rho_2^k(\mathbf{a},\mathbf{b};\mathbf{a},\mathbf{b}) = n^k(\mathbf{a})n^k(\mathbf{b}) - \mathrm{Tr}_s\,\Gamma^k(\mathbf{a};\mathbf{b})\Gamma^k(\mathbf{b};\mathbf{a}), \tag{30}$$

and the pair exchange surface (1b) generalizes to

$$w_{mn}^k(\mathbf{a},\mathbf{b}) = \chi_m^{k\dagger}(\mathbf{b})\chi_n^k(\mathbf{b})\chi_n^{k\dagger}(\mathbf{a})\chi_m^k(\mathbf{a}), \tag{31}$$

whose sum over pairs is the trace above, expressible as the squared Frobenius norm of $\Gamma^k(\mathbf{a};\mathbf{b})$. Two facts of Section 2.3 must now be re-verified, since P4 rests on them. *Zero "mass":* integrating over both arguments factorizes the surface into $\iint w_{mn}^k\, da\, db = |<\chi_m^k|\chi_n^k>|^2 = 0$ for $m \neq n$; the exchange surface is still a signed measure of zero total "mass", and the obstruction of Section 2.3, and with it the partition definition P4, remain valid without amendment. *Coincidence:* at **a=b** the pair surface is $w_{ij}^k(\mathbf{a},\mathbf{a}) = \left|\chi_i^{k\dagger}(\mathbf{a})\chi_j^k(\mathbf{a})\right|^2$, and the Cauchy–Schwarz inequality in the two-dimensional spin space gives

$$n_i^k(\mathbf{a})n_j^k(\mathbf{a}) - \left|\chi_i^{k\dagger}(\mathbf{a})\chi_j^k(\mathbf{a})\right|^2 \geq 0, \qquad n_i^k \equiv \chi_i^{k\dagger}\chi_i^k, \tag{32}$$

with equality where the local spinors are parallel. The Pauli suppression at coincidence is full for locally parallel spins, absent for locally antiparallel ones, and interpolates continuously in between; it is produced by the wave sector alone, without any dynamical input. The scalar identity (13) of Section 2.3 is its collinear limit, where the local spinors are either parallel or orthogonal.

The one-body operator of the spinor replica in a magnetic field $\mathbf{B}(\mathbf{r},t)$ carries the Zeeman term,

$$h_s = h\mathbf{1}_2 - \mu_B\,\mathbf{B}(\mathbf{r},t)\cdot\boldsymbol{\sigma}, \tag{33}$$

where $h$ is the one-body operator of Eq. (2), $\mu_B$ is the Bohr magneton, $\boldsymbol{\sigma}$ is the vector of Pauli matrices and $\mathbf{1}_2$ is the 2×2 unit matrix. The operator $h_s$ is a 2×2 Hermitian operator, local in space.

**The NW step.** The sampling assumption A0 reads: the walker $\mathbf{r}_j^l$ is distributed according to the *total spinor density* $\chi_j^{l\dagger}\chi_j^l(\mathbf{r},t)$ of its own guide spinor. Walkers carry no spin index, since they sample charge, and P4 assigns them nothing else. The prior of the Bayes step is the replica mixture of total densities,

$$n_j(\mathbf{r},t) = \frac{1}{M}\sum_{l=1}^{M}\chi_j^{l\dagger}\chi_j^l(\mathbf{r},t), \tag{34}$$

and the chain of Section 2.2, namely the conditional model (P1), the Bayes posterior and the empirical plug-in, proceeds without alteration, since at no point did it use anything about the density beyond its positivity and its normalization. The result is the identical NW potential

$$V_{\mathrm{eff}}^{k}\left(\mathbf{r}_i\right)=\sum_{j\neq i}\frac{\sum_l V_{ee}\left(\mathbf{r}_i-\mathbf{r}_j^l\right)K_{\sigma_j}\left(\mathbf{r}_j^l-\mathbf{r}_j^k\right)}{\sum_l K_{\sigma_j}\left(\mathbf{r}_j^l-\mathbf{r}_j^k\right)} \tag{35}$$

acting as a scalar times $\mathbf{1}_2$ unity matrix: it enters both component equations identically. The conditioning remains spin-blind by construction.

The graded no-double-counting condition is one of the two places where the scalar statement changes its form. At coincidence the effective pair density of the estimator reads

$$\rho_{2,\mathrm{eff}}^{ij}\left(\mathbf{a},\mathbf{a}\right)=\left[\,n_i\,n_j-\left|\chi_i^{\dagger}\chi_j\right|^2\,\right]\left(\mathbf{a}\right)+h_{ij}\left(\mathbf{a},\mathbf{a}\right), \tag{36}$$

replica indices suppressed, where $h_{ij}$ is the walker correlation hole, the difference between the stationary joint density of the walker pair of a replica, averaged over replicas, and the reference pair density built from the same guide waves. Nonnegativity of the effective pair density (36) therefore requires

$$h_{ij}\left(\mathbf{a},\mathbf{a}\right)\geq-\left[\,n_i\,n_j-\left|\chi_i^{\dagger}\chi_j\right|^2\,\right]\left(\mathbf{a}\right), \tag{37}$$

a *graded* condition: the sharp bound $h_{ij}\left(\mathbf{a},\mathbf{a}\right)\geq 0$ of the scalar theory is required only on the set where the local spinors are parallel, where the exchange cancellation is exact and the wave sector has already emptied the coincidence region, and relaxes continuously toward antiparallel alignment, where the bound (37) reaches $-n_in_j$ and the conditioning may lower the pair density by as much as the full product $n_in_j$. The collinear same-spin case saturates the parallel branch pointwise, recovering the condition of the scalar theory; the collinear opposite-spin case sits entirely on the antiparallel branch, where the condition is vacuous, consistent with the absence of exchange between opposite-spin orbitals in the collinear theory. The diagnostic changes accordingly: the same-spin pair histogram, which is the natural diagnostic of the scalar condition, becomes the *alignment-weighted* pair histogram, binned against the local weight $\left|\chi_i^{\dagger}\chi_j\right|^2/\left(n_in_j\right)$ evaluated on the guide waves at the walker pair's coincidence neighborhood. If the conditioning violates (37), the remedy is a prior built from the conditional pair density in place of the exchange-free prior of the Bayes step, and the graded inequality above is the constraint it must respect.

**Loss of the common generator, and the projected spinor equation.** After the replacement $J_i^k\rightarrow V_{\mathrm{eff}}^k\left(\mathbf{r}_i\right)$, orbital $i$ of replica $k$ evolves under its own generator

$$F_i^k=h_s+V_{\mathrm{eff}}^k\left(\mathbf{r}_i\right)-\hat{K}_i^k, \tag{38}$$

Hermitian on the spinor space but different for different $i$. Consequence (b) of Appendix A applies unchanged, and intra-replica orthogonality decays at the spinor rate derived in Appendix B.

Unlike the scalar case, no pair's orthogonality is protected by the spin structure, and the spinor extension therefore cannot use the known gauge-fixed equations. In the collinear case, opposite-spin pairs are orthogonal identically through their spin labels, and they need no multiplier; spinor orbitals are not spin eigenstates, the inner product $\chi_i^{k\dagger}\chi_j^k$ is a genuine spatial function for every pair, and every pair requires a multiplier. Step 4 of Appendix A goes through with the spinor inner product: the constraint condition determines the anti-Hermitian part of the multiplier matrix, the canonical gauge (P5) selects

$$\lambda_{ij}^{k} = < \chi_j^k \,|\, F_i^k \chi_i^k >, \quad i \neq j, \quad \lambda_{ii}^{k} = 0, \tag{39}$$

and the equation of motion of the method becomes the spinor form of (18):

$$i\hbar \partial_t \chi_i^k = F_i^k \chi_i^k - \sum_{j\neq i} \lambda_{ij}^k \chi_j^k , \tag{40}$$

with the sum over *all* $j \neq i$. P5 requires one amendment: the Löwdin realization must orthonormalize all $N$ spinor orbitals of the replica in the spinor inner product, not the same-spin subsets, since there are no same-spin subsets to orthonormalize. In imaginary time the construction applies with $-\hbar\partial_\tau$ as in Section 3; the constraint fixes $\lambda_{ji}^k + \lambda_{ij}^{k*}$ and the same formula serves for $i \neq j$, while for $i = j$ the norm is conserved with $\lambda_{ii}^k = \left\langle \chi_i^k \,\middle|\, F_i^k \chi_i^k \right>$, which is equivalent to restoring the norm after each step.

**Component form, non-collinear case. Writing the spinor equation in components**, with the shared exchange scalar $c_{ji}^k(\mathbf{r}) = \int V_{ee}(\mathbf{r}-\mathbf{r}')\chi_j^{k\dagger}\chi_i^k(\mathbf{r}')d\mathbf{r}'$ and $\Pi_{is}^k = \sum_{j\neq i} \lambda_{ij}^k \varphi_{js}^k$ the projection component:

$$i\hbar\partial_t \varphi_{i\uparrow}^k = \left(h - \mu_B B_z + V_{\text{eff}}^k\right)\varphi_{i\uparrow}^k - \mu_B\left(B_x - iB_y\right)\varphi_{i\downarrow}^k - \sum_{j\neq i} c_{ji}^k \varphi_{j\uparrow}^k - \Pi_{i\uparrow}^k , \tag{41}$$

$$i\hbar\partial_t \varphi_{i\downarrow}^k = -\mu_B\left(B_x + iB_y\right)\varphi_{i\uparrow}^k + \left(h + \mu_B B_z + V_{\text{eff}}^k\right)\varphi_{i\downarrow}^k - \sum_{j\neq i} c_{ji}^k \varphi_{j\downarrow}^k - \Pi_{i\downarrow}^k . \tag{42}$$

Four couplings are visible: the diagonal Zeeman terms split the channels energetically; the off-diagonal Zeeman terms $-\mu_B\left(B_x \mp iB_y\right)$ rotate spin through the transverse field; the exchange scalar $c_{ji}^k$ , which is one integral per pair, projected onto each component of $\chi_j^k$ , couples the components through the inter-electron spin coherence; and the projection components enforce the constraint pairwise across the whole replica. The conditional potential appears once, identically, in both lines.

**Collinear limit.** Take $B_x = B_y = 0$ and block spinors: each $\chi_m^k$ has only one nonvanishing component, $\chi_m^k = \left(\varphi_{m\uparrow}^k, 0\right)$ or $\left(0, \varphi_{m\downarrow}^k\right)$, which defines the spin labels $s_m$ of Appendix A. Then $\Gamma^k$ is block-diagonal, $\Gamma^k = \text{diag}\left(\gamma_\uparrow^k, \gamma_\downarrow^k\right)$ with $\gamma_s^k(\mathbf{r};\mathbf{r}') = \sum_{m\in s} \varphi_{ms}^k(\mathbf{r})\varphi_{ms}^{k*}(\mathbf{r}')$, and the exchange trace separates into two independent spin channels,

$$\text{Tr}_s\, \Gamma^k(\mathbf{r};\mathbf{r}')\Gamma^k(\mathbf{r}';\mathbf{r}) = \left|\gamma_\uparrow^k(\mathbf{r};\mathbf{r}')\right|^2 + \left|\gamma_\downarrow^k(\mathbf{r};\mathbf{r}')\right|^2 , \tag{43}$$

so that $E_X^k = 1/2\sum_{m\neq n} \delta_{s_m s_n} K_{mn}^k$: the $\delta_{s_m s_n}$ restriction of (A.2) is derived from the block structure, not assumed. The exchange scalar $c_{ji}^k$ vanishes for opposite-block pairs, and the block structure makes the cross-block multipliers vanish: $F_i^k$ is block-diagonal in this limit, since the one-body operator $h_s$ of Eq. (33) is diagonal in spin space, $V_{\text{eff}}^k$ is a scalar and $\hat{K}_i^k$ does not couple the blocks, so $F_i^k \chi_i^k$ stays in the block of $\chi_i^k$ and $\lambda_{ij}^k =< \chi_j^k \, |F_i^k \chi_i^k >= 0$ whenever $s_i \neq s_j$. The projection term collapses to same-spin pairs, as in the collinear formulation of Appendix A, and the spinor equation decouples into

$$i\hbar \partial_t \varphi_{is_i}^k = \left(h \mp \mu_B B_z + V_{\text{eff}}^k\right)\varphi_{is_i}^k - \sum_{j\neq i} \delta_{s_i s_j} \hat{K}_{ij}^k \varphi_{is_i}^k - \sum_{j\neq i} \delta_{s_i s_j} \lambda_{ij}^k \varphi_{js_i}^k, \tag{44}$$

the scalar equation (18) with the Zeeman splitting. The scalar theory of Sections 2.2–2.4 is therefore the collinear limit of the spinor theory. Switching on a transverse field component re-activates the cross-block multipliers and the cross-block source contributions which this limit switched off.

**Walker sector.** Both stages of Section 3 remain valid. In real time the walkers move by the guidance law of Stage 2, the velocity being the Pauli current of the total spinor density, $\mathrm{v} = \left(\hbar / m_{\mathrm{e}}\right) \mathrm{Im}\left(\chi^\dagger \nabla \chi\right) / \left(\chi^\dagger \chi\right)$, and in imaginary time they relax by the drift–diffusion process of Stage 1, the drift being the osmotic velocity of the same density, $\mathrm{u} = \left(\hbar / 2m_{\mathrm{e}}\right) \nabla \ln\left(\chi^\dagger \chi\right)$; with transverse field components the imaginary-time spinor components are in general complex, and the drift is taken from $\chi^\dagger \chi$ regardless. The same velocity field, with no separate spin variable attached to the particle, is used in the causal theory of the Pauli equation (see also [44], §9.3). The equivariance analysis of Section 3 transfers term by term. The Zeeman term is local in space and Hermitian in spin space, hence contributes nothing to $\partial_t \left(\chi^\dagger \chi\right)$: it generates no source. $V_{\text{eff}}^k$ is real, local, and scalar: no source. The exchange and projection terms generate the spinor source

$$s_i^k = -\frac{2}{\hbar} \mathrm{Im}\, \chi_i^{k\dagger} \left( \hat{K}_i^k \chi_i^k + \sum_{j\neq i} \lambda_{ij}^k \chi_j^k \right), \tag{45}$$

whose total “mass” vanishes by the same two independent mechanisms as in the scalar case, the exchange part by Hermiticity of $\hat{K}_i^k$ on the spinor space, the projection part by spinor orthonormality, and the velocity correction (26) remains solvable. It remains valid with the total spinor density in place of the scalar one, and with the projection part of the source now running over all orbital pairs. The correction is therefore not optional in this regime. It acts on the spin-summed density, which by P4 is the density the walkers sample.

## 5. Numerical examples

The spinor equations of Section 4 are tested by a simple example of two electrons in a one-dimensional double quantum dot, the elementary unit of spin qubits [56–58], in a magnetic field with a transverse gradient such as a micromagnet produces [59]. The distance between the dots tunes the exchange coupling from strongly coupled dots in which the electrons share an orbital to separated dots holding one electron each, and the gradient turns the quantization axis from one dot to the other, so that the spins are not collinear and all spin-dependent terms of Eqs. (40)–(42) are active. We compare with the numerically exact solution on the same grid and consider only the imaginary-time stage, the preparation of the ground state.

### 5.1 Model and methods

The confining potential is the one-dimensional cut along the axis of the quartic double dot (e.g. in [60]),

$$V(x) = \frac{\omega^2}{8a^2}\left(x^2 - a^2\right)^2, \tag{46}$$

where $x$ is the coordinate along the axis and $\omega$ the frequency of each well; the minima lie at $x = \pm a$, so that $a$ is half the distance between the dots. In atomic units the two-electron Hamiltonian is

$$H = \sum_{i=1}^{2}\left[-\frac{1}{2}\frac{\partial^2}{\partial x_i^2} + V(x_i) - \mu_{\mathrm{B}}\mathbf{B}(x_i)\cdot\boldsymbol{\sigma}_i\right] + \left(1 + (x_1 - x_2)^2\right)^{-1/2}, \tag{47}$$

where $x_i$ is the coordinate of electron $i$, $\mu_B = 1/2$, and the last term the soft-Coulomb electron-electron interaction. The field is

$$\mathbf{B}(x) = B_0\mathbf{e}_z + \beta x\mathbf{e}_x. \tag{48}$$

Here $\mathbf{e}_x$ and $\mathbf{e}_z$ are the unit vectors along $x$ and $z$, $B_0$ is a uniform field and $\beta$ the gradient of the transverse component. In one dimension the field enters only through the Zeeman term, while the spin keeps its full rotational structure, as in nanowire double dots [61]. The fields at the centers of the two dots, $\mathbf{B}(\pm a) = (\pm\beta a, 0, B_0)$, enclose the angle

$$\theta_B = 2\arctan\frac{\beta a}{B_0}, \tag{49}$$

The methods used for numerical solutions are presented in Appendix C, and the derivations of the scalar and the spinor equations in Appendices A and B. We compare two forms of the TDQMC replica. In the first, each replica is a single determinant of two spinor orbitals propagated by Eqs. (40)–(42). In the second, each replica is the spin-coupled pair of valence-bond theory, Eq. (C.2), built from two non-orthogonal spatial orbitals and a $2\times 2$ spin matrix, so that a single replica can hold any real superposition of the singlet and the three triplet states. The determinant is the special case of the pair in which the spin matrix has rank one. In both forms the walkers carry no spin. The determinant form is the method derived in Sections 2–4 and Appendices A and B. The spin-coupled form is an extension which we introduce for the crossover regime, and only its imaginary-time equations are given (Appendix C.3), obtained from the energy of the pair with the direct interaction conditioned as in Section 2.2.

### 5.2 Results

Figure 1 shows the von Neumann entropy $S - \ln 2$ and the angle between the two local spins $\theta_{\mathrm{LR}}$ against $a$ where three regimes appear.

**Strong exchange,** $a \le 1.5$**.** The exact state is nearly a singlet ($\theta_{\mathrm{LR}} \ge 178.8°$, $P_{\mathrm{S}} \ge 0.995$), and its entanglement, which comes from the spatial correlation of the electrons, grows from 0.019 at $a = 0.5$ to 0.245 at $a = 1.5$. HF gives none as expected. Both forms of TDQMC recover 40%–91% of it, for both determinant and spin-coupled replicas at $a = 0.5$, 1.0, 1.5. For the spin-coupled pairs most of it is carried by the pair itself, and the conditional interaction adds the rest, 0.015 at $a = 1.5$. At $a = 1.5$ the determinant replicas, like HF, already tilt the spins, $\theta_{\mathrm{LR}} = 172.7°$, while the spin-coupled replicas give the exact angle. At $a = 1.5$ the singlet probability (C.9) is 0.996 for the exact state and the spin-coupled replicas and 0.854 for the determinant replicas.

**Crossover,** $a = 2.0$–$2.25$**.** Of the distances computed, the exact entanglement is largest at $a = 2.0$ ($S - \ln 2 = 0.300$, $\theta_{\mathrm{LR}} = 157.6°$). Eq. (C.12) gives 0.267 and 0.050 at $a = 2.0$ and 2.25 against the exact 0.300 and 0.050: the field cuts the zero-field branch $S_0(a)$ where $J$ falls below $2\beta a$, and a stronger gradient cuts it earlier, where the electrons are less localized, and lowers the peak; at $a = 2.0$ the exact $S - \ln 2$ is 0.431, 0.300, 0.150 for $\beta = 0.005$, 0.01, 0.02, and Eq. (C.12) gives 0.405, 0.267, 0.134 (the gradients 0.005 and 0.02 are not shown). Here the two replica forms differ.

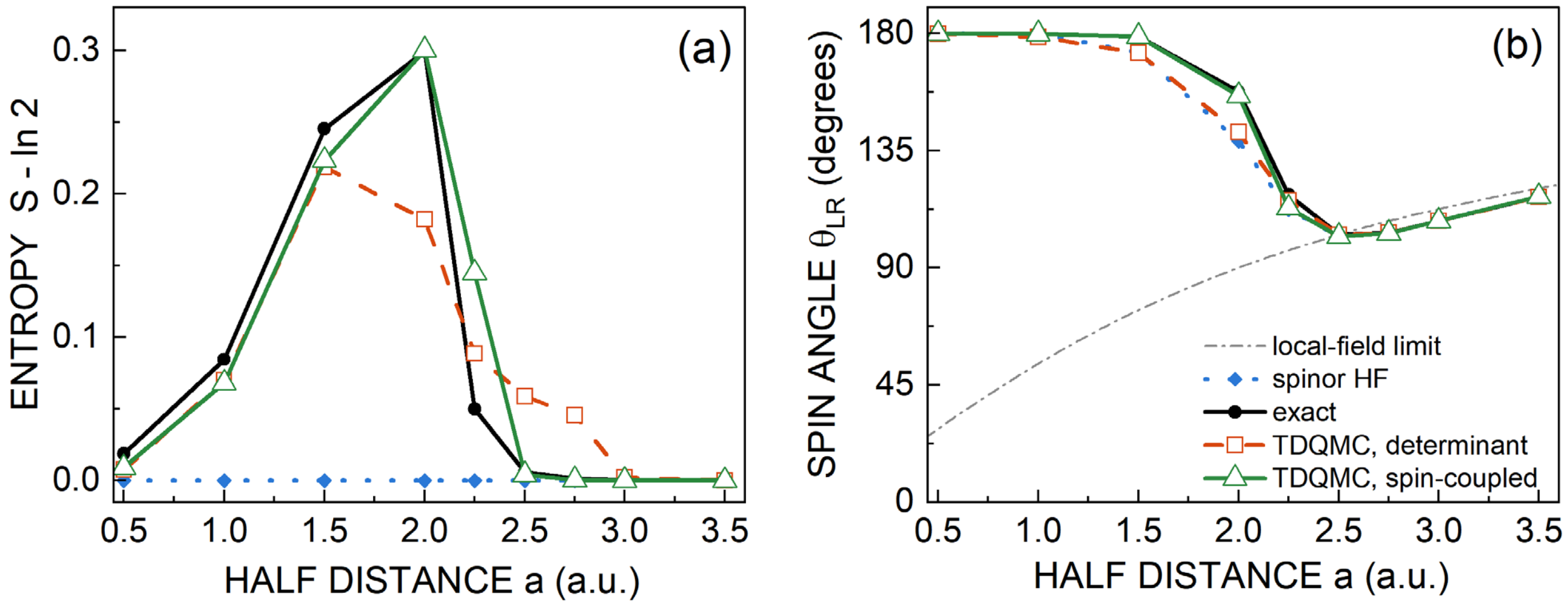

**Fig. 1.** Two electrons in the double dot (46)–(48) with $\omega = 1$, $B_0 = 0.02$ and $\beta = 0.01$ at the end of the imaginary-time stage, against the half distance $a$ between the wells: (a) one-body entanglement entropy in excess of a single determinant, Eq. (C.8); (b) angle between the spins of the two dots, Eq. (C.7). Exact (black lines, circles), spinor HF (blue dot lines, diamonds), and TDQMC with $\sigma_{1,2} = 1$ a.u., whose replicas are single determinants, Eqs. (40)–(42) (red dashed lines, squares), or spin-coupled pairs, Eq. (C.2) [57,62–64] (green lines, triangles). Spinor HF is the mean-field baseline, to which determinant TDQMC reduces for $\sigma_{1,2} \to \infty$. Dash-dot line in (b): the local-field limit $\theta_B$, Eq. (49), the angle between two independent spins aligned with the fields at the two minima.

Determinant TDQMC gives $S - \ln 2 = 0.182$, $\theta_{\mathrm{LR}} = 142.1°$, close to HF (138.7°): by Eq. (C.10) a determinant of localized orbitals is at most half singlet when its orbitals do not overlap, and the conditional potential acts on positions, not spins, so each replica carries the singlet–triplet superposition only at mean-field level. The spin-coupled pair (C.2) carries it, giving $S - \ln 2 = 0.300$

, $\theta_{\mathrm{LR}} = 156.2°$. The singlet probability (C.9) confirms this, 0.818 for the spin-coupled pairs and 0.556 for the determinant replicas against the exact 0.828 at $a = 2.0$, and each spin-coupled replica exceeds the largest singlet probability of a single determinant of its orbitals (Appendix C.2). The spin-coupled pairs give the geometry already without the conditional interaction, which adds spatial correlation, and over four random sequences their $S - \ln 2$ at $a = 2.0$ is $0.33 \pm 0.03$. At $a = 2.25$ all angles are within 6° of the exact 118.1°, but both forms overestimate the entanglement, 0.089 (determinant) and 0.145 (spin-coupled; 0.19 with another random sequence) against 0.050.

**Separated dots,** $a \geq 2.5$**.** The exact angle joins the local-field limit at $a = 2.5$, where the entanglement has dropped to 0.006: the excess of $\theta_{\mathrm{LR}}$ over $\theta_B$, 67.6° at $a = 2.0$ and 21.4° at $a = 2.25$, falls with the entanglement, and $\theta_{\mathrm{LR}}$ passes through its minimum of 102.7° where the two meet. For $a \geq 2.75$ the exact angle lies 3°–5° below $\theta_B$, a one-body effect: each spin follows the field at the centroid of its electron, which lies inside the minimum because the well is softer towards the barrier, and at the centroids of the HF orbitals the field angle is 108.0° and 117.4° at $a = 3.0$ and 3.5, against the exact 108.0° and 117.3°. Both TDQMC forms reproduce the exact angle within 0.6°, angles no collinear calculation can give. The determinant replicas, however, overestimate the small entanglement there. Splitting $S$ into the entropy $S_{\mathrm{sp}}$ of the spin-traced matrix and the remainder $S - S_{\mathrm{sp}}$ (Appendix C.2) shows that the spin-coupled replicas reproduce both parts, whereas the determinant replicas reach $S - \ln 2$ at strong exchange with too large a spatial and too small a spin part, and their excess at separated dots is a spread of spin directions between the replicas, which builds up while the walkers are fixed.

The model is one-dimensional, without spin–orbit coupling [52,53]; only the imaginary-time stage is treated with $\sigma_{1,2}$ held at 1 a.u. For determinant replicas the real-time equations are those of Section 4; for the spin-coupled pair (C.2) the real-time guidance of the walkers remains to be formulated. Within these limits, spinor TDQMC still reproduces the non-collinear spin geometry, which no collinear calculation can represent, and recovers 40%–91% of the one-body entanglement entropy where the exchange is strong, for determinant replicas with a deficient spin part. In the crossover, where the coherence between singlet and triplet determines the geometry, this coherence must be carried by the replicas themselves, and the spin-coupled pair (C.2) is the minimal replica which carries it, as its singlet probability confirms.

## 6. Summary

In this work we have constructed the conditional interaction of time-dependent quantum Monte Carlo. Taking the replica mixture of the guide-wave densities as the prior and the walker as a measurement of the electron position with finite resolution, we obtained from Bayes' theorem and a single empirical substitution the Nadaraya–Watson form which the method had been using heuristically. For each walker the posterior is exact for the conditional model, and the construction is approximate Bayesian in the sense of the established approximate Bayesian computation, becoming exact Bayes as the width vanishes. The nonlocal length $\sigma_j$ is then the width of a conditional distribution and is fixed in general by minimizing the energy (P2), and the pair limit and the mean-field limit are obtained for zero and for infinite width.

For fermions we find three consequences, none of which is visible in the singlet benchmarks. The exchange surface is a signed measure of zero total "mass", which admits no positive, normalized walker representation; it therefore remains in the wave sector, and the walker dynamics stays positive and in physical space. The conditioning destroys the common generator of the mean-field theory, so the multipliers of the orthonormality constraint can no longer be removed; we derive the projection term which replaces them, fixed only up to a Hermitian addition. The exchange and projection terms break the equivariance of the walker guidance through a source of zero total "mass", and this vanishing "mass" is the solvability condition of the correction which we derive. We generalize the construction to spinors by two substitutions. The Pauli suppression at coincidence then depends on the local spin alignment, and the spinor theory reduces through the mean-field limit to the known two-particle spin equations and through the collinear limit to the scalar theory.

The advantages of the representation differ between the two statistics. For bosons a replica is a Hartree product, with no exchange operator, no orthonormality constraint, no projection term and no equivariance defect. The method then consists only of the conditional potential and the walker dynamics. For fermions the determinant is kept in each replica and the exchange is carried by the Fock operator. The Pauli suppression at coincidence is then reproduced replica by replica at the level of the determinant, while the walkers remain a positive, exchange-free ensemble. In both cases the sampling is in physical space rather than in configuration space, the number of coupled one-body equations grows linearly with the number of particles, and the cost is polynomial in the number of electrons, of replicas and of grid points. The method is formulated in real time, which distinguishes it from the projector techniques, although the calculations of this work are confined to imaginary time.

The double-dot example of Section 5 tests the spinor equations in imaginary time, where the spinor TDQMC reproduces the non-collinear spin geometry of two electrons in an inhomogeneous field and recovers between 40% and 91% of the exact one-body entanglement entropy for strong exchange coupling, determinant replicas through a spatial excess, and spin-coupled replicas in both parts. In the crossover between the exchange and the Zeeman regimes, determinant replicas carry the spin coupling only at the mean-field level, while replicas in the form of the spin-coupled pair of valence-bond theory carry it already without the conditional interaction and give the exact spin angle and singlet probability to within 6° and 0.03. The conditional interaction adds spatial correlation, towards the exact entanglement where the exchange is strong and beyond it in the crossover.

Several extensions follow from this work. The spinor equations are intended for the real-time, exchange-driven dynamics of spatially overlapping electrons whose spinors are not collinear: gate-controlled spin rotation, exchange-mediated entanglement between spin qubits, and singlet–triplet mixing driven by an inhomogeneous field. Because the ensemble consists of one-body objects labeled by walker positions, entanglement and coherence acquire a spatial argument and can be mapped in real space at the cost of the dynamics itself. For bosons the same equations apply without the exchange, projection and source terms, and condensate dynamics is therefore the simplest case in which the conditional interaction can be studied separately. The nonlocal length $\sigma_j$ and the screening length $r_s$ are in general position dependent; here they are held constant as a first approximation.

## Appendix A. Derivation of the scalar TDQMC equations

The TDHF equation is normally written in gauge-fixed form [3,31], where the Lagrange multipliers of the orthonormality constraint are removed by a unitary transformation. That removal is valid only while all orbitals of a replica evolve under one common Hermitian generator. Since the NW replacement destroys this property, the constraint has to be carried through the replacement explicitly. We do this in four steps: the constrained action (Step 1), its stationarity conditions (Step 2), the two consequences of the replacement (Step 3), and the determination of the multipliers (Step 4).

**Step 1: The constrained action of a replica.** Each replica $k$ carries orbitals $\varphi_1^k,\ldots,\varphi_N^k$ with fixed spin labels $s_1,\ldots,s_N$ [65], and $h=-\dfrac{\hbar^2}{2m_e}\nabla^2+V_{en}(\mathbf{r})$ denotes the one-body operator. The pair energies are written from the outset in the self-interaction-free convention,

$$E_H^k=\frac{1}{2}\sum_{m\neq n}J_{mn}^k,\qquad J_{mn}^k=\iint V_{ee}(\mathbf{r}-\mathbf{r}')\left|\varphi_m^k(\mathbf{r})\right|^2\left|\varphi_n^k(\mathbf{r}')\right|^2 d\mathbf{r}\,d\mathbf{r}', \tag{A.1}$$

$$E_X^k=-\frac{1}{2}\sum_{m\neq n}\delta_{s_m s_n}K_{mn}^k,\qquad K_{mn}^k=\iint V_{ee}(\mathbf{r}-\mathbf{r}')\varphi_m^{k*}(\mathbf{r})\varphi_n^k(\mathbf{r})\varphi_n^{k*}(\mathbf{r}')\varphi_m^k(\mathbf{r}')d\mathbf{r}\,d\mathbf{r}'. \tag{A.2}$$

Step 3 shows why this convention is required. The action of the replica, with the orthonormality constraints adjoined, is

$$S^k=\int dt\left[\sum_i<\varphi_i^k\,|\,i\hbar\partial_t-h\,|\varphi_i^k>-E_H^k-E_X^k+\sum_{i,j}\lambda_{ij}^k\left(<\varphi_i^k\,|\,\varphi_j^k>-\delta_{ij}\right)\right]. \tag{A.3}$$

The quantities $\lambda_{ij}^k$ are Lagrange multipliers, one for each orthonormality constraint $<\varphi_i^k\,|\,\varphi_j^k>=\delta_{ij}$, and they are fixed by the requirement that the constraints continue to hold as the orbitals evolve. For the constraint term to be real, the multiplier matrix at the action level must be Hermitian, $\lambda_{ij}^{k*}=\lambda_{ji}^k$. Hermitian multipliers are sufficient at the TDHF level but not after the NW replacement, and the constraint must then be imposed on the equations of motion instead of the action.

**Step 2: Stationarity, term by term.** Vary with respect to $\varphi_i^{k*}(\mathbf{r})$ and set $\delta S^k=0$. The one-body part gives $(i\hbar\partial_t-h)\varphi_i^k(\mathbf{r})$ directly. For the Hartree energy, $\varphi_i^{k*}$ appears in (A.1) in the terms with $m=i$ and, by the symmetry $J_{mn}=J_{nm}$, equally in those with $n=i$; the two families are equal and cancel the prefactor $1/2$:

$$\frac{\delta E_H^k}{\delta\varphi_i^{k*}(\mathbf{r})}=\left[\sum_{j\neq i}\int V_{ee}(\mathbf{r}-\mathbf{r}')\left|\varphi_j^k(\mathbf{r}')\right|^2 d\mathbf{r}'\right]\varphi_i^k(\mathbf{r})\equiv J_i^k\,\varphi_i^k(\mathbf{r}). \tag{A.4}$$

For the exchange energy, $\varphi_i^{k*}$ appears at the first argument of the $m=i$ terms and at the second argument of the $n=i$ terms; again the two contributions are equal and cancel the $1/2$:

$$\frac{\delta E_X^k}{\delta\varphi_i^{k*}(\mathbf{r})} = -\sum_{j\neq i}\delta_{s_i s_j}\,\varphi_j^k(\mathbf{r})\int V_{ee}(\mathbf{r}-\mathbf{r}')\varphi_j^{k*}(\mathbf{r}')\varphi_i^k(\mathbf{r}')\,d\mathbf{r}' \equiv -\hat{K}_i^k\,\varphi_i^k(\mathbf{r}). \tag{A.5}$$

Equation (A.5) defines the exchange operator used throughout: $\hat{K}_i^k$ is the total, spin-contracted Fock term acting on orbital $i$ of replica $k$, $\hat{K}_i^k = \sum_{j\neq i}\delta_{s_i s_j}\hat{K}_{ij}^k$, where $\hat{K}_{ij}^k\varphi_i^k(\mathbf{r}) = \varphi_j^k(\mathbf{r})\int V_{ee}(\mathbf{r}-\mathbf{r}')\varphi_j^{k*}(\mathbf{r}')\varphi_i^k(\mathbf{r}')\,d\mathbf{r}'$ is the exchange operator of the pair. The constraint term contributes $\sum_j \lambda_{ij}^k\varphi_j^k(\mathbf{r})$. Collecting,

$$i\hbar\,\partial_t\varphi_i^k = \left(h + J_i^k\right)\varphi_i^k - \hat{K}_i^k\,\varphi_i^k - \sum_j \lambda_{ij}^k\,\varphi_j^k. \tag{A.6}$$

At the TDHF level the multipliers can be removed. We give the argument, since it fails after the NW replacement. First, an on-shell identity: the excluded self-terms are harmless, since the direct and exchange self-contributions acting on the own orbital coincide pointwise,

$$\left[\int V_{\mathrm{ee}}(\mathbf{r}-\mathbf{r}')\left|\varphi_i^k(\mathbf{r}')\right|^2 d\mathbf{r}'\right]\varphi_i^k(\mathbf{r}) - \varphi_i^k(\mathbf{r})\int V_{\mathrm{ee}}(\mathbf{r}-\mathbf{r}')\varphi_i^{k*}(\mathbf{r}')\varphi_i^k(\mathbf{r}')\,d\mathbf{r}' = 0, \tag{A.7}$$

so the restricted sums in (A.4)–(A.5) may be freely extended to full sums when acting on $\varphi_i^k$. In spin-orbital form the extended operator $F^k = h + \sum_j\left(J_j^k - \hat{K}_j^k\right)$ is then *one common Hermitian generator for all orbitals of the replica* (the factor $\delta_{s_m s_n}$ arising automatically from the spin contraction). With a common Hermitian generator, orthonormality is preserved already at $\lambda = 0$, since $\frac{d}{dt} < \varphi_i^k \mid \varphi_j^k > = (i\hbar)^{-1}\left(< \varphi_i^k \middle| F^k\varphi_j^k > - < F^k\varphi_i^k \middle| \varphi_j^k >\right) = 0$ by Hermiticity; and any Hermitian $\lambda^k$ one chooses to retain generates only a time-dependent unitary mixing of the orbitals, under which the determinant is invariant, so it may be transformed away. This gauge fixing rests on the identity (A.7) and on the common generator.

**Step 3: The NW replacement and its two consequences.** P1–P3 replace the Hartree operator of orbital $i$ by the conditional potential built from the walkers of the *other* electrons,

$$J_i^k \to V_{\mathrm{eff}}^k(\mathbf{r}_i) = \sum_{j\neq i}\frac{\sum_l V_{\mathrm{ee}}\left(\mathbf{r}_i - \mathbf{r}_j^l\right)K_{\sigma_j}\left(\mathbf{r}_j^l - \mathbf{r}_j^k\right)}{\sum_l K_{\sigma_j}\left(\mathbf{r}_j^l - \mathbf{r}_j^k\right)}. \tag{A.8}$$

*Consequence (a): the self-interaction-free convention is forced.* Suppose the full-sum convention had been used in (A.1), relying on the cancellation (A.7). After the replacement, the direct self-term would be the NW-smoothed potential of the orbital's own walker, at small $\sigma_j$ essentially $V_{\mathrm{ee}}\left(\mathbf{r}-\mathbf{r}_i^k\right)\varphi_i^k(\mathbf{r})$, while the exchange self-term remains the exact integral on the left of (A.7). These are different objects; the cancellation (A.7) is destroyed and the equation acquires a spurious self-interaction of the order of their difference. The $j \neq i$ convention of (A.1)–(A.2) removes both self-terms identically *before* the replacement and is therefore the only consistent starting point.

*Consequence (b): the common generator is lost.* After the replacement, orbital $i$ of replica $k$ evolves under

$$F_i^k = h + V_{\text{eff}}^k(\mathbf{r};i) - \hat{K}_i^k, \tag{A.9}$$

and $F_i^k \neq F_j^k$: the conditional potentials of different electrons are built from different walker sets and differ (not only by on-shell-vanishing terms) once the walkers separate. Setting $\lambda = 0$ in (A.6) then gives, for $i \neq j$, using the Hermiticity of each individual $F_i^k$,

$$\frac{d}{dt} < \varphi_i^k \mid \varphi_j^k >= \frac{1}{i\hbar} < \varphi_i^k \mid \left(F_j^k - F_i^k\right) \mid \varphi_j^k > \neq 0: \tag{A.10}$$

intra-replica orthogonality in general decays at a rate set by the matrix element of the *difference* of conditional potentials (plus the difference of exchange operators). The unitary gauge argument of Step 2 is now unavailable, since a unitary transformation mixes orbitals obeying equations with *different* generators and cannot absorb the constraint forces. The multipliers discarded at the TDHF level must be reinstated. The exchange energy (A.2) and the per-replica bound $E^k \geq E_{\text{MF}}$, the mean-field minimum, both presuppose orthonormal orbitals, and fail as (A.10) accumulates; this happens in the parallel-spin cases, but not for opposite-spin two-electron systems, where orthogonality is carried by the spin labels.

**Step 4: Determining the multipliers; the final equations.** Reinstate the constraint term and write the trial equation of motion

$$i\hbar\partial_t \varphi_i^k = F_i^k \varphi_i^k - \sum_j \lambda_{ij}^k \varphi_j^k. \tag{A.11}$$

Impose that orthonormality, assumed at time $t$, is preserved. Differentiating, substituting (A.11) twice, once for $\partial_t \varphi_j^k$ and once, in conjugated form using the Hermiticity of $F_i^k$, for $\partial_t \varphi_i^k$, and using $< \varphi_i^k \mid \varphi_m^k >= \delta_{im}$ to collapse the multiplier sums, we obtain

$$\frac{d}{dt} < \varphi_i^k \mid \varphi_j^k >= \frac{1}{i\hbar}\left[ < \varphi_i^k \left| F_j^k \varphi_j^k \right> - \lambda_{ji}^k - < \varphi_i^k \left| F_i^k \right| \varphi_j^k > + \lambda_{ij}^{k*} \right]. \tag{A.12}$$

Setting this to zero for all pairs gives the constraint condition

$$\lambda_{ji}^k - \lambda_{ij}^{k*} = < \varphi_i^k \left| F_j^k \varphi_j^k \right> - < \varphi_i^k \left| F_i^k \right| \varphi_j^k >. \tag{A.13}$$

Two facts follow from (A.13). First, a Hermitian multiplier matrix has $\lambda_{ji}^k - \lambda_{ij}^{k*} = 0$ and can satisfy (A.13) only if the right-hand side vanishes, which is the TDHF case of a common generator. After the NW replacement the right-hand side is in general nonzero, so *the action-level (Hermitian) multipliers of (A.3) cannot enforce the constraint*; the constraint force must be imposed at the level of the equations of motion, where $\lambda^k$ need not be Hermitian. This is also why the projected equation below is not the stationarity condition of any action: the resulting constraint-preserving force is not the functional derivative of the TDQMC energy functional. Second, (A.13) fixes only the combination $\lambda_{ji}^k - \lambda_{ij}^{k*}$; adding any Hermitian matrix to $\lambda^k$ preserves orthonormality and constitutes a residual orbital-mixing gauge freedom.

The canonical (minimal) choice takes the diagonal to vanish, $\lambda_{ii}^{k}=0$, which preserves each norm, and for $i\neq j$

$$\lambda_{ij}^{k}=<\varphi_{j}^{k}\,|\,F_{i}^{k}\,\varphi_{i}^{k}>, \tag{A.14}$$

which satisfies (A.13) identically: $\lambda_{ji}^{k}=<\varphi_{i}^{k}\,|\,F_{j}^{k}\varphi_{j}^{k}>$ and $\lambda_{ij}^{k*}=<\varphi_{j}^{k}\left|F_{i}^{k}\varphi_{i}^{k}\right>^{*}=<\varphi_{i}^{k}\left|F_{i}^{k}\,\right|\varphi_{j}^{k}>$ by Hermiticity of $F_{i}^{k}$, and their difference reproduces the right-hand side of (A.13). Substituting into (A.11) yields the **exchange-involved TDQMC equation of motion** in orthonormality-preserving form:

$$i\hbar\,\partial_{t}\varphi_{i}^{k}=F_{i}^{k}\,\varphi_{i}^{k}-\sum_{j\neq i}\varphi_{j}^{k}<\varphi_{j}^{k}\,|\,F_{i}^{k}\,\varphi_{i}^{k}>=\left(1-\sum_{j\neq i}\left|\varphi_{j}^{k}\right><\varphi_{j}^{k}\right|\right)F_{i}^{k}\,\varphi_{i}^{k}, \tag{A.15}$$

with $F_{i}^{k}$ given by (17): each orbital is driven by its own generator, with the components along the other orbitals of the same replica projected out. Equation (A.15) is Eq. (18) of the main article, the equation of motion of the method. In the collinear (scalar) formulation the constraint binds only within same-spin blocks, since opposite-spin orbitals are orthogonal through their spin labels regardless of spatial overlap, so that the projection sum runs over same-spin $j$ only; in the spinor formulation the transverse components of the field mix the blocks and the sum runs over all $j\neq i$.

In (18) the direct electron–electron Coulomb interaction *is* $V_{\text{eff}}^{k}\left(\mathbf{r}_{i}\right)$, and the bare $V_{\text{ee}}$ also appears as the kernel inside the exchange integral, but that is the coherence sector. The Coulomb interaction then enters a second time, through the multipliers: $\lambda_{ij}^{k}=<\varphi_{j}^{k}\,|\,F_{i}^{k}\varphi_{i}^{k}>$ contains matrix elements of $V_{\text{eff}}^{k}\left(\mathbf{r}_{i}\right)$, and by (A.10) the orthogonality decay is driven chiefly by the difference $V_{\text{eff}}^{k}\left(\mathbf{r}_{j}\right)-V_{\text{eff}}^{k}\left(\mathbf{r}_{i}\right)$. The conditional Coulomb term that generates the correlation therefore also sets the size of the projection.

The reduction to the mean field requires two conditions. As $\sigma_{j}\to\infty$ the kernel weights become uniform and the posterior ceases to condition, so that by (9) the conditional potential becomes the Hartree potential of the mixture density, the same in every replica; with $M\to\infty$ taken after $\sigma_{j}\to\infty$ the replicas coincide and $n_{j}=\left|\varphi_{j}^{k}\right|^{2}$. The generator loses its replica index in this limit, but it keeps its orbital index, since $V_{\text{eff}}^{k}$ is built from the electrons other than $i$. The self-interaction-free convention removes what remains: by the identity (A.7) the direct and the exchange self-terms cancel when acting on $\varphi_{i}^{k}$, so that the restricted sums may be extended to full sums and the generator (17) becomes one common Hermitian Fock operator $F^{k}$ for every orbital of the replica.

With a common generator the multiplier matrix is Hermitian, $\lambda_{ij}^{*}=\lambda_{ji}$, and the orbital mixing $U$ which solves $i\hbar\,\partial_{t}U=-\lambda U$ is therefore unitary. It leaves the one-body density matrix invariant, and with it $F^{k}$ itself, and it multiplies the determinant by $\det U$, which is a phase. The projection term is therefore transformed away and (18) reduces to the time-dependent Hartree–Fock equation (2). At finite $\sigma_{j}$ the construction fails at its first step, since $\lambda_{ji}-\lambda_{ij}^{*}=<\varphi_{i}^{k}\left|\left(F_{j}^{k}-F_{i}^{k}\right)\right|\varphi_{j}^{k}>$ does not vanish:

the mixing is no longer unitary, and the anti-Hermitian part of the multiplier matrix, which is the constraint force, cannot be removed by any choice of gauge.

The kinetic energy can be taken in either of two forms:

$$E_{\mathrm{kin}}^{k}=\frac{\hbar^{2}}{2m_{\mathrm{e}}}\sum_{i}\int\left|\nabla\varphi_{i}^{k}\right|^{2}d\mathbf{r}\ \text{or}\ E_{\mathrm{kin}}^{k}=-\frac{\hbar^{2}}{2m_{\mathrm{e}}}\sum_{i}\left(\frac{\nabla^{2}\varphi_{i}^{k}}{\varphi_{i}^{k}}\right)_{\mathbf{r}_{i}^{k}} \tag{A.16}$$

the first evaluated on the grid and the second at the walker. Writing the one-body term this way leaves the choice open in the estimator itself.

The TDQMC energy estimator (P3) is

$$E=\frac{1}{M}\sum_{k}\left(E_{\mathrm{kin}}^{k}+\sum_{i}V_{\mathrm{en}}\left(\mathbf{r}_{i}^{k}\right)-\frac{1}{2}\sum_{i\neq j}\delta_{s_{i}s_{j}}K_{ij}^{k}+\sum_{i>j}V_{\mathrm{ee}}\left(\mathbf{r}_{i}^{k}-\mathbf{r}_{j}^{k}\right)\right). \tag{A.17}$$

The one-body terms are written here at the walker, as in the published TDQMC calculations. They may equally be taken as grid quantities of the guide waves, which carry no walker variance; under A0 the two agree in expectation, but that equality is untested, since the published calculations use the walker form and the grid form has been exercised only at the Hartree–Fock level. Their difference in a correlated run therefore measures the systematic deviation of the walkers from their own guide densities, and since both follow from a single run it is a diagnostic of A0 at no additional cost.

The interaction term admits two readouts of $E\left(\sigma_{j}\right)$, and both are used in practice. The particle estimator is (A.17), which takes both electrons of each pair at their walkers. The mixed walker–wave estimator holds one electron of each pair at its walker and traces the other over that partner's own guide density,

$$E=\frac{1}{M}\sum_{k}\left(E_{\mathrm{kin}}^{k}+\sum_{i}V_{\mathrm{en}}\left(\mathbf{r}_{i}^{k}\right)-\frac{1}{2}\sum_{i\neq j}\delta_{s_{i}s_{j}}K_{ij}^{k}+\sum_{i>j}\int V_{\mathrm{ee}}\left(\mathbf{r}_{j}^{k}-\mathbf{r}\right)\left|\varphi_{i}^{k}\left(\mathbf{r}\right)\right|^{2}d\mathbf{r}\right). \tag{A.18}$$

In both, the one-body and exchange contributions are grid quantities of the guide waves evaluated with the bare interaction, since by P3 the conditional potential enters neither. The second has the structure of a Rao–Blackwellization [66,67] of the first: it agrees with it in expectation to the extent that the guide density of the traced electron reproduces its conditional density given the retained walker, with equality as $\sigma_{j}\to\infty$, and it is the preferable readout for parallel spins, where weaker orbital overlap makes the bare pair sum noisy.

It can be shown that for $N$ electrons and $M$ replicas on a grid of $G$ points per guide wave the numerical load is proportional to $NMG\left(N\log G+M\right)$, the partner potentials of the conditional interaction being formed once per replica, and the memory to $NMG$. This is polynomial in $N$, $M$ and $G$, against the $G^{N}$ of the many-body wave function on the same grid [41], and the calculation costs about $M$ times a Hartree–Fock calculation. The decomposition is moreover parallel by construction: within a time step the replicas are coupled only through the pooled walker positions, so that distributing them leaves a single reduction per step as the only communication, although the guide waves inside a replica remain coupled through the exchange operator and the orthonormality projection.

## Appendix B. Derivation of the spinor TDQMC equations

The spinor construction of Section 4 is derived here in the four steps of Appendix A. We write the steps out because every inner product changes meaning: $<u\,|\,v>=\int u^{\dagger}v(\mathbf{r})d\mathbf{r}$ now contracts the two spinor components as well as integrating over space, and each orbital is a two-component field.

**Step 1: The constrained action of a replica.** Each replica $k$ carries $N$ spinor orbitals $\chi_1^k,\dots,\chi_N^k$, orthonormal in the spinor inner product, $<\chi_i^k\,|\,\chi_j^k>=\delta_{ij}$, and the one-body operator is the $h_s$ of (33), which carries the Zeeman term. The pair energies, in the self-interaction-free convention forced by Step 3 of Appendix A, are

$$E_H^k=\frac{1}{2}\sum_{m\neq n}J_{mn}^k,\quad J_{mn}^k=\iint V_{ee}(\mathbf{r}-\mathbf{r}')\chi_m^{k\dagger}\chi_m^k(\mathbf{r})\chi_n^{k\dagger}\chi_n^k(\mathbf{r}')d\mathbf{r}\,d\mathbf{r}', \tag{B.1}$$

$$E_X^k=-\frac{1}{2}\sum_{m\neq n}K_{mn}^k,\quad K_{mn}^k=\iint V_{ee}(\mathbf{r}-\mathbf{r}')\chi_m^{k\dagger}(\mathbf{r})\chi_n^k(\mathbf{r})\chi_n^{k\dagger}(\mathbf{r}')\chi_m^k(\mathbf{r}')d\mathbf{r}\,d\mathbf{r}'. \tag{B.2}$$

The Hartree energy is manifestly spin-blind, since it couples total charge densities. The exchange energy carries no $\delta_{s_m s_n}$ restriction: the spinor inner products include all spin combinations automatically, and the restriction emerges in the collinear limit. The action of the replica, with the orthonormality constraints adjoined, is

$$S^k=\int dt\left[\sum_i<\chi_i^k\left|(i\hbar\partial_t-h_s)\chi_i^k\right\rangle-E_H^k-E_X^k+\sum_{i,j}\lambda_{ij}^k\left(\left\langle\chi_i^k\right|\chi_j^k>-\delta_{ij}\right)\right]. \tag{B.3}$$

The quantities $\lambda_{ij}^k$ are Lagrange multipliers, one for each constraint $<\chi_i^k\,|\,\chi_j^k>=\delta_{ij}$. Reality of the constraint term requires the multiplier matrix to be Hermitian at the action level, as in Appendix A.

**Step 2: Stationarity, term by term.** Vary with respect to $\chi_i^{k\dagger}(\mathbf{r})$ and set $\delta S^k=0$. The one-body part gives $(i\hbar\partial_t-h_s)\chi_i^k$ directly. In the Hartree energy $\chi_i^{k\dagger}$ appears at the first argument of the $m=i$ terms and at the second argument of the $n=i$ terms; the two contributions coincide after relabeling and cancel the factor $\frac{1}{2}$, leaving

$$J_i^k\chi_i^k(\mathbf{r})=\sum_{j\neq i}\int V_{ee}(\mathbf{r}-\mathbf{r}')\chi_j^{k\dagger}\chi_j^k(\mathbf{r}')d\mathbf{r}'\,\chi_i^k(\mathbf{r}), \tag{B.4}$$

a real scalar multiple of the identity in spin space, which therefore acts identically on both components of $\chi_i^k$. The same counting in the exchange energy gives the spinor Fock operator

$$\hat{K}_i^k \chi_i^k(\mathbf{r}) = \sum_{j\neq i} \chi_j^k(\mathbf{r}) \int V_{ee}(\mathbf{r}-\mathbf{r}') \chi_j^{k\dagger}(\mathbf{r}') \chi_i^k(\mathbf{r}') d\mathbf{r}', \tag{B.5}$$

in which the integral is a complex scalar, namely the full spinor inner product of the two orbitals under the Coulomb kernel, multiplying the two-component spinor $\chi_j^k(\mathbf{r})$. The operator therefore couples both spin components of orbital $i$ through a single shared integral, and no Kronecker restriction appears, since there is no spin label. Collecting the three contributions together with the constraint term,

$$i\hbar \partial_t \chi_i^k = \left(h_s + J_i^k\right) \chi_i^k - \hat{K}_i^k \chi_i^k - \sum_j \lambda_{ij}^k \chi_j^k. \tag{B.6}$$

The on-shell self-term identity of Appendix A survives for spinors. Setting $j = i$ in the kernels of (B.4) and (B.5),

$$\int V_{ee}(\mathbf{r}-\mathbf{r}') \chi_i^{k\dagger} \chi_i^k(\mathbf{r}') d\mathbf{r}' \chi_i^k(\mathbf{r}) - \chi_i^k(\mathbf{r}) \int V_{ee}(\mathbf{r}-\mathbf{r}') \chi_i^{k\dagger}(\mathbf{r}') \chi_i^k(\mathbf{r}') d\mathbf{r}' = 0, \tag{B.7}$$

since each term is the real scalar $\int V_{ee} \chi_i^{k\dagger} \chi_i^k(\mathbf{r}') d\mathbf{r}'$ multiplying $\chi_i^k(\mathbf{r})$. The restricted sums in (B.4) and (B.5) may therefore be extended to full sums when acting on $\chi_i^k$, and consequence (a) of Step 3, namely that the self-interaction-free convention is forced once the NW replacement is made, holds for spinors by the argument of Appendix A unchanged.

**Step 3: The NW replacement.** P1–P3 replace the Hartree operator $J_i^k$ by the conditional potential (35), built from the walkers of the other electrons of the replica. That kernel is real and local, so $V_{\text{eff}}^k(\mathbf{r};i)$ enters as a multiple of the identity in spin space and introduces no spin structure of its own. The generator of orbital $i$ becomes

$$F_i^k = h_s + V_{\text{eff}}^k(\mathbf{r};i) - \hat{K}_i^k, \tag{B.8}$$

Hermitian on the spinor space, but different for different $i$, since the conditional potentials of different electrons are built from different walker sets. The common generator of Step 2 is lost, and setting $\lambda^k = 0$ in (B.6) then gives, for $i \neq j$ and using the Hermiticity of each $F_i^k$,

$$\frac{d}{dt} < \chi_i^k \middle| \chi_j^k >= \frac{1}{i\hbar} < \chi_i^k \middle| F_j^k - F_i^k \,|\, \chi_j^k > \neq 0, \tag{B.9}$$

so that intra-replica orthogonality in general decays at a rate set by the matrix element of the difference of the conditional potentials, together with the difference of the exchange operators. The unitary gauge argument is unavailable for the same reason as in the scalar case, and the multipliers discarded at the spinor TDHF level must be reinstated.

**Step 4: Determining the multipliers.** Reinstate the constraint term and write the trial equation of motion

$$i\hbar \partial_t \chi_i^k = F_i^k \chi_i^k - \sum_j \lambda_{ij}^k \chi_j^k. \tag{B.10}$$

Impose that orthonormality, assumed at time $t$, is preserved. Differentiate the overlap of two orbitals of the replica, substitute the trial equation twice, once for $\partial_t \chi_j^k$ and once, conjugated, for $\partial_t \chi_i^k$, and use $<\chi_i^k \mid \chi_m^k>=\delta_{im}$ to collapse the multiplier sums:

$$\frac{d}{dt}<\chi_i^k \mid \chi_j^k>=\frac{1}{i\hbar}\left(<\chi_i^k\left|F_j^k\,\chi_j^k\right>-\lambda_{ji}^k-<\chi_i^k\left|F_i^k\,\chi_j^k\right>+\lambda_{ij}^{k*}\right). \tag{B.11}$$

Setting this to zero for every pair gives the constraint condition

$$\lambda_{ji}^k-\lambda_{ij}^{k*}=<\chi_i^k\left|F_j^k\,\chi_j^k\right>-<\chi_i^k\left|F_i^k\,\chi_j^k\right>. \tag{B.12}$$

The right-hand side is in general nonzero once the NW replacement is made, so, as in Appendix A, a Hermitian multiplier matrix cannot satisfy (B.12) and the constraint force must be imposed at the level of the equations of motion. The canonical choice (39) satisfies (B.12) identically, by the Hermiticity of each $F_i^k$, and substituting it into the trial equation returns the projected spinor equation (40). No pair is protected by a spin label here, so the condition is active for every $i \neq j$; the $\delta_{s_i s_j}$ restriction of the scalar theory reappears only in the collinear limit, as shown at the end of Section 4.

The estimators (A.17) and (A.18) remain valid with $n_i^k=\chi_i^{k\dagger}\chi_i^k$, the one-body, exchange and Zeeman terms being grid integrals of the guide spinors. The mixed form (A.18), which is evaluated during the propagation in the calculations, lies between (A.17) and the per-replica Hartree integral $J_{ij}^k$ of (A.1), to which it collapses when the retained walker is traced as well, so that the difference between (A.17) and (A.18) measures the two-body correlation which the one-sided average misses.

The spinor theory is the top of a chain of limits:

spinor TDQMC → spinor TDHF → two-particle spin equations (4a-4d) in [60],

spinor TDQMC → collinear (scalar) TDQMC → scalar TDHF,

the arrows of the first line being the mean-field limit $\sigma_j \to \infty$ and then $N=2, M=1$ at zero vector potential, and those of the second the collinear reduction, block spinors with $B_x=B_y=0$, followed by the same mean-field limit. The two reductions are independent and they commute, and the scalar theory is the collinear limit of the spinor theory at every resolution length. The two-particle spin equations sit on the upper line: they are its mean-field limit for two electrons and a single replica, with a general field and both spinor components populated, so they describe the non-collinear case restricted to a single determinant.

## Appendix C. Methods for numerical solutions

### C.1 Numerical parameters

For the numerical examples we use $\omega=1$, $B_0=0.02$, $\beta=0.01$, nine half distances, $a=0.5$, 1.0, 1.5, 2.0, 2.25, 2.5, 2.75, 3.0 and 3.5, and a grid of 128 points spanning 20 a.u. (10 a.u. at $a=0.5$), and the time step $\Delta\tau=0.05$. The spins are coupled by the singlet–triplet splitting $J=E_{\mathrm{T}}-E_{\mathrm{S}}$ of the exact lowest singlet and triplet energies at zero field, the exchange coupling of the spin-qubit literature [56,57], not the Fock exchange. $J$ falls from 0.43 a.u. at $a=1.0$ to 0.040 at $a=2.0$ and $2.2\times10^{-4}$ at $a=3.0$, and near $a=2$ it crosses the field difference $2\beta a$ of the dots: the scan passes from a locked singlet to two spins each following its local field.

**Exact solution.** The ground state is the limit $\tau\to\infty$ of the imaginary-time equation for the two-electron spinor $\Psi_{s_1s_2}(x_1,x_2)$ with the spin indices $s_1,s_2\in\{\uparrow,\downarrow\}$,

$$-\frac{\partial\Psi_{s_1s_2}}{\partial\tau}=\left[h(x_1)+h(x_2)+\left(1+(x_1-x_2)^2\right)^{-1/2}\right]\Psi_{s_1s_2}-\mu_\mathrm{B}\sum_{s'}\left[\left(\mathbf{B}(x_1)\cdot\boldsymbol{\sigma}\right)_{s_1s'}\Psi_{s's_2}+\left(\mathbf{B}(x_2)\cdot\boldsymbol{\sigma}\right)_{s_2s'}\Psi_{s_1s'}\right],\text{(C.1)}$$

where $h(x)=-\frac{1}{2}\frac{\partial^2}{\partial x^2}+V(x)$, $(\boldsymbol{B}\cdot\boldsymbol{\sigma})_{ss'}$ is the matrix element of $\boldsymbol{B}\cdot\boldsymbol{\sigma}$ between the spin states $s,s'\in\{\uparrow,\downarrow\}$, and $\Psi_{s_1s_2}(x_1,x_2)=-\Psi_{s_2s_1}(x_2,x_1)$. It is integrated by the split-step Fourier method with antisymmetrization after each step, up to $\tau=250$. Spinor (generalized) Hartree–Fock (HF) [54,55] is the single-determinant limit of the same Hamiltonian, propagated in imaginary time. At each $a$ all calculations start from the in-field HF ground state; for TDQMC its two orbitals are first localized by a Foster–Boys rotation [68]. The in-field HF ground state has two broken-symmetry solutions, A and its mirror image B, which differ by exchanging the spin states of the left and the right orbital. The determinant replicas $k=1,\ldots,M/2$ start from the localized orbitals of A and the replicas $k=M/2+1,\ldots,M$ from those of B. All spin-coupled pairs start from A, with $u^k=|\chi_1|$, $w^k=|\chi_2|$ and $C^k_{s_1s_2}=a_{s_1}b_{s_2}$, where $\chi_1$ and $\chi_2$ are the localized HF spinors and $a_s$, $b_s$ their normalized spin vectors, so that each pair starts as the HF determinant and imaginary time relaxes $C^k$ from there.

**TDQMC.** We compare two forms of the replica, both with $M=256$ replicas, the conditional potential (35) with the nonlocal length $\sigma_{1,2}=1$ a.u., and $\tau=500$. In the first, replica $k$ is the determinant of two spinor orbitals $\chi_1^k$, $\chi_2^k$ propagated by Eqs. (40)–(42). Since the attachment of a walker to an orbital, unlike the determinant, changes under a unitary mixing of the two orbitals, the orbitals are rotated after every exchange step into the pair with the most separated centroids [68]. This rotation leaves the determinant unchanged and is a choice of the multiplier gauge of P5 (Section 2.4), and like that choice it is part of the specification of the calculation, since it changes the attachment of the walkers to the orbitals. The walkers are sampled by the Metropolis algorithm in the first pass, $\tau\le25$, and then held fixed. We verify A0 through the replica averages of $x^2$ over the walkers and over the guide densities, the latter being the second moment of $n_j$. After the first pass this moment of $n_j$ changes by less than 0.5% for $a\ge2.0$ and by at most 3.7% for $a\le1.5$, so that the fixed walkers sample the final density. For $a\ge2.0$ their second moment exceeds that of $n_j$ by 4–9%, and for $a\le1.5$ it differs by up to 15%, at most about twice the standard deviation, 2.6–7.7%, with which the second moment of the 256 walkers fluctuates during the first pass, and the difference is therefore the sampling error of a single set of walkers. All distances use the same random sequence. During the passes with fixed walkers the energy decreases by at most 1.5%, while $S-\ln2$ changes by up to 0.05 (Section 5.2). In the second form, replica $k$ is the spin-coupled pair of valence-bond theory [57,62–64],

$$\Psi^k_{s_1 s_2}\left(x_1,x_2\right)=u^k\left(x_1\right)w^k\left(x_2\right)\mathbf{C}^k_{s_1 s_2}-w^k\left(x_1\right)u^k\left(x_2\right)\mathbf{C}^k_{s_2 s_1}, \tag{C.2}$$

with real, normalized, non-orthogonal orbitals $u^k$ (electron 1) and $w^k$ (electron 2) and a real $2\times 2$ spin matrix $\boldsymbol{C}^k$, whose element $C^k_{s_1 s_2}$, $s_1,s_2\in\{\uparrow,\downarrow\}$, is the amplitude with which the electron in $u^k$ has spin $s_1$ and the electron in $w^k$ has spin $s_2$, normalized as $\sum_{s_1 s_2}|C^k_{s_1 s_2}|^2=1$. Its antisymmetric part gives the singlet amplitude $\left(C^k_{\uparrow\downarrow}-C^k_{\downarrow\uparrow}\right)/\sqrt{2}$ and its symmetric part the three triplet amplitudes, so that one replica holds any real superposition of the four two-spin states; for $\boldsymbol{C}^k$ of rank one, (C.2) is a single determinant. The norm of the pair is $<\Psi^k|\Psi^k>=2\left[1-<u^k|w^k>^2\sum_{s_1 s_2}C^k_{s_1 s_2}C^k_{s_2 s_1}\right]$, and the energy and $\gamma$ of each replica are evaluated with $\Psi^k$ divided by this norm. In each replica, $u^k$, $w^k$ and $\mathbf{C}^k$ relax in imaginary time towards the minimum of the energy of $\Psi^k$ for the Hamiltonian (47), with the direct interaction of each electron replaced by the conditional potential (35) of the other electron's walker and the exchange-type terms taken from the replica's own orbitals. The equations are given in Appendix C.3. The two walkers sample $\left|u^k\right|^2$ and $\left|w^k\right|^2$ by the Metropolis algorithm in every pass. In both forms the walkers carry no spin; all spin correlation therefore resides in the replicas. The determinant results change by less than 0.5° in $\theta_{\mathrm{LR}}$ and 0.01 in $S-\ln 2$ over the last 100 a.u.; with walkers resampled in every pass, the spin-coupled results fluctuate by up to 0.7° and 0.02, most at $a=2.25$. Between random sequences the spin-coupled $S-\ln 2$ at $a=2.0$ spreads more, from 0.300 to 0.365 (Section 5.2). For $\sigma_{1,2}=10^{10}$ a.u., where (35) becomes the Hartree potential, determinant TDQMC reproduces HF to within 1°. Notice that (C.2) is the two-electron case of a replica built as an antisymmetrized product of spin-coupled pairs; its $N$-electron extension is polynomial when spin correlation is confined within pairs, and full $N$-electron spin coupling is less affordable.

### C.2 Observables

All quantities derive from the spin-resolved one-body density matrix, a $2\times 2$ matrix in spin space at each pair of points $x,x'$. For a two-electron state $\Psi$ its elements are

$$\gamma_{ss'}\left(x;x'\right)=2\sum_{s_2}\int\Psi_{ss_2}\left(x,x_2\right)\Psi^*_{s's_2}\left(x',x_2\right)dx_2, \tag{C.3}$$

with $\mathrm{Tr}\gamma=2$. For TDQMC, Eq. (C.3) evaluated with $\Psi^k$ gives the matrix $\Gamma^k$ of replica $k$, for a determinant $\Gamma^k\left(x;x'\right)=\sum_i\chi^k_i\left(x\right)\chi^{k\dagger}_i\left(x'\right)$ as in Eq. (29), and

$$\gamma\left(x;x'\right)=\frac{1}{M}\sum_{k=1}^{M}\Gamma^k\left(x;x'\right). \tag{C.4}$$

The charge and spin densities are

$$n\left(x\right)=\mathrm{Tr}_s\gamma\left(x;x\right),\qquad \mathbf{s}\left(x\right)=\mathrm{Tr}_s\left[\boldsymbol{\sigma}\gamma\left(x;x\right)\right], \tag{C.5}$$

with $\mathrm{Tr}_s$ the trace over spin, and the spin polarization of each dot is

$$\mathbf{m}_\mathrm{L} = \frac{\int_{x<0} \mathbf{s}(x)\,dx}{\int_{x<0} n(x)\,dx}, \qquad \mathbf{m}_\mathrm{R} = \frac{\int_{x>0} \mathbf{s}(x)\,dx}{\int_{x>0} n(x)\,dx}. \tag{C.6}$$

The angle between the spins of the two dots,

$$\theta_\mathrm{LR} = \arccos\frac{\mathbf{m}_\mathrm{L}\cdot\mathbf{m}_\mathrm{R}}{|\mathbf{m}_\mathrm{L}||\mathbf{m}_\mathrm{R}|}, \tag{C.7}$$

can only be 0 or 180° in a collinear calculation. The entanglement is the von Neumann entropy

$$S = -\mathrm{Tr}\left[\frac{\gamma}{2}\ln\frac{\gamma}{2}\right], \tag{C.8}$$

with the trace over space and spin. For two fermions $S \geq \ln 2$, with equality for a single determinant, so that $S - \ln 2$ vanishes for HF; in TDQMC it arises from the differences between the replicas [28–30] and, for spin-coupled replicas, also within each replica. For the exact state, which is pure, $S - \ln 2$ measures the entanglement in excess of that of a single determinant, and we call it the one-body entanglement entropy. The matrix (C.3) of a pure state is an average over the position of the partner electron, $\gamma(x,x') = \int n(x_2)\rho_{x_2}(x,x')\,dx_2$, and TDQMC forms (C.4) in the same way, the orbitals of replica $k$ being conditioned on the walkers of their partners (P1), which sample the density (A0). In determinant replicas, however, the spin part of each conditional state is pure, and a mixture of determinants can have the same $\gamma$ as an entangled state, so we also use the split of $S$ into the entropy $S_\mathrm{sp}$ of the spin-traced matrix, which is $\ln 2$ for any determinant of orbitals localized in different dots, and $S - S_\mathrm{sp}$, together with the singlet probability. The singlet probability

$$P_S = \mathrm{Tr}(\Pi_S\rho_2) = \frac{1 - \mathrm{Tr}(\boldsymbol{\sigma}_1\cdot\boldsymbol{\sigma}_2\rho_2)}{4}, \tag{C.9}$$

where $\Pi_\mathrm{S}$ projects onto the singlet $\left(|\uparrow\downarrow\rangle - |\downarrow\uparrow\rangle\right)/\sqrt{2}$ and $\rho_2$ is $|\Psi><\Psi|$ for the exact state and $M^{-1}\sum_k |\Psi^k><\Psi^k|$ for TDQMC, separates the singlet from the triplet, which for separated dots have the same $\gamma$. It therefore tests the replica ensemble at the two-particle level, which the one-body entropy (C.8) does not. For a determinant with orbitals localized in different dots and unit spin directions $\mathbf{n}_\mathrm{L}$, $\mathbf{n}_\mathrm{R}$,

$$P_S = \frac{1 - \mathbf{n}_\mathrm{L}\cdot\mathbf{n}_\mathrm{R}}{4} = \frac{1}{2}\sin^2\frac{\theta_\mathrm{LR}}{2} \leq \frac{1}{2}, \tag{C.10}$$

so that it is at most half singlet. For orbitals with the overlap $\Omega$ the bound becomes $P_\mathrm{S} \leq (1+\Omega^2)/2$, which is reached for antiparallel spins. The separated dots behave as two spins coupled by $J$ in the local fields,

$$H_\mathrm{spin} = \frac{J}{4}\boldsymbol{\sigma}_1\cdot\boldsymbol{\sigma}_2 - \mu_\mathrm{B}(\mathbf{B}_\mathrm{L}\cdot\boldsymbol{\sigma}_1 + \mathbf{B}_\mathrm{R}\cdot\boldsymbol{\sigma}_2), \tag{C.11}$$

with $\mathbf{B}_{\mathrm{L,R}} = \mathbf{B}(\mp a)$, whose ground state passes from the singlet at large $J$ to the product of local-field states at small $J$, in which each spin aligns with the field in its own dot. We call this the *local-field limit*. In it $\theta_{\mathrm{LR}} = \theta_B$ and $S - \ln 2 = 0$, so that the field geometry alone fixes the spins, and a departure of $\theta_{\mathrm{LR}}$ from $\theta_B$, like the entanglement, measures the exchange coupling still acting between them. At $a = 2.0$ the model gives $\theta_{\mathrm{LR}} = 158.0°$ against the exact 157.6°. Let $S_{\mathrm{spin}}(a) = -\mathrm{Tr}\rho_{\mathrm{L}} \ln \rho_{\mathrm{L}}$ be the entanglement of this ground state, $\rho_{\mathrm{L}}$ the reduced density matrix of the left spin, and $S_0(a)$ the exact $S - \ln 2$ at zero field, which rises from 0.085 at $a = 1$ towards $\ln 2$ as the electrons localize in separate dots. The estimate

$$S - \ln 2 \approx S_0(a) \frac{S_{\mathrm{spin}}(a)}{\ln 2}, \tag{C.12}$$

is exact at zero field and for fully localized orbitals, and has no adjustable parameter.

### C.3 Imaginary-time equations of the spin-coupled pair

For one replica (index $k$ dropped) let $h = -1/2\partial_x^2 + V(x)$ be the spin-free one-body operator, $B_\alpha(x)$, $\alpha \in \{x, z\}$, the components of the field (48), and for two orbitals $f, g \in \{u, w\}$ let $h_{fg} = \langle f|h|g\rangle$ and $B_\alpha^{fg} = \langle f|B_\alpha|g\rangle$. With the overlap $\Omega = \langle u|w\rangle$, the spin exchange $\eta = \mathrm{Tr}(CC) = \sum_{s_1 s_2} C_{s_1 s_2} C_{s_2 s_1}$, the local spins $m_\alpha^{(1)} = \mathrm{Tr}\left(C^T \hat{\sigma}_\alpha C\right)$, $m_\alpha^{(2)} = \mathrm{Tr}\left(C \hat{\sigma}_\alpha C^T\right)$, and the exchange spins $n_\alpha^{(1)} = \mathrm{Tr}\left(C^T \hat{\sigma}_\alpha C^T\right)$, $n_\alpha^{(2)} = \mathrm{Tr}\left(C \hat{\sigma}_\alpha C\right)$, the energy of the normalized pair (C.2) for the Hamiltonian (47) is

$$E = \frac{D - X}{1 - \Omega^2 \eta},$$

$$D = h_{uu} + h_{ww} + J_{uw} - \mu_B \sum_\alpha \left( B_\alpha^{uu} m_\alpha^{(1)} + B_\alpha^{ww} m_\alpha^{(2)} \right),$$

$$X = \eta\left(2\Omega\, h_{uw} + K_{uw}\right) - \mu_B\, \Omega \sum_\alpha B_\alpha^{uw} \left( n_\alpha^{(1)} + n_\alpha^{(2)} \right),$$

where $J_{uw} = \iint u(x)^2\, v(x-y)\, w(y)^2\, dx\, dy$ is the direct and $K_{uw} = \iint u(x) w(x) v(x-y) u(y) w(y)\, dx\, dy$ the exchange integral, with $v(x) = (1+x^2)^{-1/2}$. The orbitals follow the imaginary-time gradient of $E$, with the direct potential of the partner orbital replaced by the conditional potential (35) of its walker, and the spin matrix $C$ follows the same gradient at fixed orbitals. In $E$ the direct term is the orbital integral $J_{uw}$, so that only the orbital equations are conditioned. After each step $u$, $w$ and $C$ are normalized, and the orbitals are propagated by the split-step Fourier method. With $V_{\mathrm{eff}}^k$ replaced by the Hartree potential of $w$, these are the equations of the variational spin-coupled pair.

**Acknowledgment:** This research is based upon work supported by the Bulgarian Ministry of Education and Science as a part of National Roadmap for Research Infrastructure, grant D01-102/26.06.2025 (ELI ERIC BG), and by the National Science Fund, grant КП-06-Н78/6. The funding sources had no involvement in the study design, the calculations, the interpretation of the results, the writing of the article or the decision to submit it.

## References

[1] M. Troyer and U.-J. Wiese, Phys. Rev. Lett. 94, 170201 (2005).

[2] A. Montina, Phys. Rev. A 77, 022104 (2008).

[3] C. Froese Fischer, *The Hartree–Fock Method for Atoms: A Numerical Approach* (Wiley-Interscience, New York, 1977).

[4] P. Hohenberg and W. Kohn, Phys. Rev. 136, B864 (1964).

[5] W. Kohn and L. J. Sham, Phys. Rev. 140, A1133 (1965).

[6] E. Runge and E. K. U. Gross, Phys. Rev. Lett. 52, 997 (1984).

[7] N. T. Maitra, K. Burke, and C. Woodward, Phys. Rev. Lett. 89, 023002 (2002).

[8] R. J. Bartlett and M. Musiał, Rev. Mod. Phys. 79, 291 (2007).

[9] H.-D. Meyer, U. Manthe, and L. S. Cederbaum, Chem. Phys. Lett. 165, 73 (1990).

[10] M. H. Beck, A. Jäckle, G. A. Worth, and H.-D. Meyer, Phys. Rep. 324, 1 (2000).

[11] J. Zanghellini, M. Kitzler, C. Fabian, T. Brabec, and A. Scrinzi, Laser Phys. 13, 1064 (2003).

[12] U. Schollwöck, Rev. Mod. Phys. 77, 259 (2005).

[13] N. Metropolis, A. W. Rosenbluth, M. N. Rosenbluth, A. H. Teller, and E. Teller, J. Chem. Phys. 21, 1087 (1953).

[14] W. L. McMillan, Phys. Rev. 138, A442 (1965).

[15] M. H. Kalos, D. Levesque, and L. Verlet, Phys. Rev. A 9, 2178 (1974).

[16] D. M. Ceperley and B. J. Alder, Phys. Rev. Lett. 45, 566 (1980).

[17] P. J. Reynolds, D. M. Ceperley, B. J. Alder, and W. A. Lester, Jr., J. Chem. Phys. 77, 5593 (1982).

[18] G. Sugiyama and S. E. Koonin, Ann. Phys. (N.Y.) 168, 1 (1986).

[19] S. Zhang and H. Krakauer, Phys. Rev. Lett. 90, 136401 (2003).

[20] E. Gull, A. J. Millis, A. I. Lichtenstein, A. N. Rubtsov, M. Troyer, and P. Werner, Rev. Mod. Phys. 83, 349 (2011).

[21] B. L. Hammond, W. A. Lester, Jr., and P. J. Reynolds, *Monte Carlo Methods in Ab Initio Quantum Chemistry* (World Scientific, Singapore, 1994).

[22] W. M. C. Foulkes, L. Mitas, R. J. Needs, and G. Rajagopal, Rev. Mod. Phys. 73, 33 (2001).

[23] B. M. Austin, D. Y. Zubarev, and W. A. Lester, Jr., Chem. Rev. 112, 263 (2012).

[24] F. Krausz, Rev. Mod. Phys. 96, 030502 (2024).

[25] I. P. Christov, Opt. Express 14, 6906 (2006).

[26] I. P. Christov, New J. Phys. 9, 70 (2007).

[27] I. P. Christov, J. Opt. Soc. Am. B 34, 1817 (2017).

[28] I. P. Christov, Entropy 25, 1308 (2023).

[29] I. P. Christov, Entropy 27, 1093 (2025).

[30] I. P. Christov, ACS Omega 11, 4837 (2026).

[31] P. A. M. Dirac, Proc. Cambridge Philos. Soc. 26, 376 (1930).

[32] J. Frenkel, *Wave Mechanics: Advanced General Theory* (Clarendon, Oxford, 1934).

[33] A. Gelman, J. B. Carlin, H. S. Stern, D. B. Dunson, A. Vehtari, and D. B. Rubin, *Bayesian Data Analysis*, 3rd ed. (CRC Press, Boca Raton, 2013).

[34] R. D. Wilkinson, Stat. Appl. Genet. Mol. Biol. 12, 129 (2013).

[35] J.-M. Marin, P. Pudlo, C. P. Robert, and R. J. Ryder, Stat. Comput. 22, 1167 (2012).

[36] E. A. Nadaraya, Theory Probab. Appl. 9, 141 (1964).

[37] G. S. Watson, Sankhyā Ser. A 26, 359 (1964).

[38] I. P. Christov, Few-Body Syst. 61, 45 (2020).

[39] I. P. Christov, Entropy 23, 868 (2021).

[40] I. P. Christov, J. Chem. Phys. 136, 034116 (2012).

[41] I. P. Christov, J. Phys. Chem. A 113, 6016 (2009).

[42] P.-O. Löwdin, J. Chem. Phys. 18, 365 (1950).

[43] D. Bohm, Phys. Rev. 85, 166 (1952); 85, 180 (1952).

[44] P. R. Holland, *The Quantum Theory of Motion* (Cambridge University Press, Cambridge, 1993).

[45] E. Nelson, *Quantum Fluctuations* (Princeton University Press, Princeton, 1985).

[46] G. Parisi and Y. Wu, Sci. Sin. 24, 483 (1981).

[47] M. Pavon, J. Math. Phys. 37, 3375 (1996).

[48] I. P. Christov, J. Chem. Phys. 127, 134110 (2007).

[49] G. O. Roberts and R. L. Tweedie, Bernoulli 2, 341 (1996).

[50] I. P. Christov, J. Chem. Phys. 128, 244106 (2008).

[51] T. A. Elsayed, K. Mølmer, and L. B. Madsen, Sci. Rep. 8, 12704 (2018).

[52] Y. A. Bychkov and E. I. Rashba, J. Phys. C 17, 6039 (1984).

[53] G. Dresselhaus, Phys. Rev. 100, 580 (1955).

[54] H. Fukutome, Int. J. Quantum Chem. 20, 955 (1981).

[55] J. Kübler, K.-H. Höck, J. Sticht, and A. R. Williams, J. Phys. F 18, 469 (1988).

[56] D. Loss and D. P. DiVincenzo, Phys. Rev. A 57, 120 (1998).

[57] G. Burkard, D. Loss, and D. P. DiVincenzo, Phys. Rev. B 59, 2070 (1999).

[58] R. Hanson, L. P. Kouwenhoven, J. R. Petta, S. Tarucha, and L. M. K. Vandersypen, Rev. Mod. Phys. 79, 1217 (2007).

[59] M. Pioro-Ladrière, T. Obata, Y. Tokura, Y.-S. Shin, T. Kubo, K. Yoshida, T. Taniyama, and S. Tarucha, Nat. Phys. 4, 776 (2008).

[60] D. M. Sullivan and D. S. Citrin, J. Appl. Phys. 96, 1540 (2004).

[61] S. Nadj-Perge, S. M. Frolov, E. P. A. M. Bakkers, and L. P. Kouwenhoven, Nature 468, 1084 (2010).

[62] W. Heitler and F. London, Z. Phys. 44, 455 (1927).

[63] W. A. Goddard III, Phys. Rev. 157, 81 (1967).

[64] D. L. Cooper, J. Gerratt, and M. Raimondi, Chem. Rev. 91, 929 (1991).

[65] J. A. Pople and R. K. Nesbet, J. Chem. Phys. 22, 571 (1954).

[66] D. Blackwell, Ann. Math. Stat. 18, 105 (1947).

[67] G. Casella and C. P. Robert, Biometrika 83, 81 (1996).

[68] J. M. Foster and S. F. Boys, Rev. Mod. Phys. 32, 300 (1960).